\documentclass[twocolumn,showpacs,amsmath,amssymb,superscriptaddress,floatfix,nofootinbib]{revtex4-1}
\usepackage{float}
\usepackage{booktabs,makecell,multirow}
\usepackage{longtable}
\usepackage{subfigure}
\usepackage{color}
\usepackage{graphicx,amsfonts}
\usepackage{amsmath}
\usepackage[dvipdfm,colorlinks=true,citecolor=blue,linkcolor=blue,urlcolor=blue]{hyperref}
\usepackage{url}
\usepackage{ulem}

\newcommand*{\RMN}[1]{\uppercase\expandafter{\romannumeral#1}}

\begin{document}

\title{Reheating constraints on modified single-field Natural Inflation models}
\author{Hua Zhou}
\email{zhouhua@cqu.edu.cn}
\address{Department of Physics, Chongqing Key Laboratory for Strongly Coupled Physics, Chongqing University, Chongqing 401331, People's
Republic of China}
\address{Department of Physics, Norwegian University of Science and Technology, H{\o}gskoleringen 5, N-7491 Trondheim, Norway}
\author{Qing Yu}
\email{yuq@cqu.edu.cn}
\address{Department of Physics, Chongqing Key Laboratory for Strongly Coupled Physics, Chongqing University, Chongqing 401331, People's
Republic of China}
\address{Department of Physics, Norwegian University of Science and Technology, H{\o}gskoleringen 5, N-7491 Trondheim, Norway}
\author{Yu Pan}
\email{panyu@cqupt.edu.cn}
\address{School of Science, Chongqing University of Posts and Telecommunications, Chongqing 400065, P.R. China}
\author{Ruiyu Zhou}
\email{zhoury@cqupt.edu.cn}
\address{School of Science, Chongqing University of Posts and Telecommunications, Chongqing 400065, P.R. China}
\author{Wei Cheng}
\email{chengwei@cqupt.edu.cn(corresponding author)}
\address{School of Science, Chongqing University of Posts and Telecommunications, Chongqing 400065, P.R. China}

\date{\today}

\begin{abstract}
In this paper, we discuss three modified single-field natural inflation models in detail, including Special generalized Natural Inflation model(SNI), Extended Natural Inflation model(ENI) and Natural Inflation inspired model(NII). We derive the analytical expression of the tensor-to-scalar ratio $r$ and the spectral index $n_s$ for those models. Then the reheating temperature $T_{re}$ and reheating duration $N_{re}$ are analytically derived. Moreover, considering the CMB constraints, the feasible space of the SNI model in $(n_s, r)$ plane is almost covered by that of the NII, which means the NII is more general than the SNI. In addition, there is no overlapping space between the ENI and the other two models in $(n_s, r)$ plane, which indicates that the ENI and the other two models exclude each other, and more accurate experiments can verify them. Furthermore, the reheating brings tighter constraints to the inflation models, but they still work for a different reheating universe. Considering the constraints of $n_s$, $r$, $N_k$ and choosing $T_{re}$ near the electroweak energy scale, one can find that the decay constants of the three models have no overlapping area and the effective equations of state $\omega_{re}$ should be within $\frac{1}{4}\lesssim \omega_{re} \lesssim \frac{4}{5}$ for the three models.
\end{abstract}
\maketitle

\section{INTRODUCTION}
The inflation theory is one of the accepted solutions to the problem of horizon and flatness in Big Bang cosmology \cite{Starobinsky:1980te,Baumann:2009ds,Kinney:2003xf,Guth:1980zm,Sato:1980yn,Albrecht:1982wi,Linde:1981mu}. The quantum fluctuations of the inflaton field provide a piece of fundamental knowledge for studying the anisotropy of the cosmic microwave background(CMB) \cite{Kolb:1990vq,Lyth:1998xn,Armendariz-Picon:1999hyi,Mukhanov:1990me,DeFelice:2010aj,Cheng:2020ocy} and the structure of the universe \cite{Guth:1982ec,Guth:1985ya,Starobinsky:1982ee,Mukhanov:1981xt}. At present, the predictions of scale-invariant inflation, Gaussian and adiabatic density perturbations have been confirmed by WMAP \cite{WMAP:2010qai}, COBE \cite{COBE:1992gfs}, Planck \cite{Planck:2018jri} and so on.

The single scalar field inflation model that relies on the slow-rolling is now the mainstream inflation model \cite{Creminelli:2014oaa,Kaloper:2008fb,Dalianis:2021iig,Steer:2003yu,Geshnizjani:2003cn,Chen:2006xjb,Seery:2005wm}, which is described by its potential $V(\phi)$. When the slope and curvature of $V(\phi)$ are small enough to satisfy the slow-roll conditions, the universe will continue to inflate. At the end of inflation, the universe goes into the next period which is usually called reheating \cite{Allahverdi:2010xz}. During the reheating, the energy density in the inflaton is transformed into a thermal bath, which fills the universe at the beginning of the era of radiation dominance. The reheating scenarios have a complex physics, where the duration of the reheating would be affected by the speed and type of particles, and there is usually a so-called preheating stage. In this stage, the inflation field decays into massive particles through non-perturbative processes such as parametric resonance and instantaneous preheating \cite{Kofman:1994rk,Kofman:1997yn,Felder:1998vq,Dufaux:2006ee}. After the preheating, the frequency band with parametric resonance will have very high occupancy, while the rest of the space will be in a highly non-thermal state \cite{Cook:2015vqa}.

The reheating stage can be parameterized with the reheating temperature $T_{re}$, the effective equation of state(eos) $\omega_{re}$ of the matter in the reheating process, and the reheating duration, i.e., number of e-foldings $N_{re}$. For the value of $T_{re}$, it should be larger than the electroweak scale to meet the requirement of producing weak-scale dark matter. In addition, to reach the temperature of big bang nucleosynthesis, $T_{re}$ should be greater than 10 MeV \cite{Kawasaki:1999na,Kawasaki:2000en}. Furthermore, considering the constraints of the late entropy produced by the decay of massive particles, the $T_{re}$ would be as low as $[2.5-4]$ MeV \cite{Kawasaki:1999na,Kawasaki:2000en}.

Reheating is an extremely complex physical process \cite{Khlebnikov:1996wr}, and it is difficult for us to directly detect and study. Typically, to avoid the complexity of reheating and simplify the description, the default choice of EOS $\omega_{re}$ is the constant in the interval $[-1/3,1]$ \cite{Podolsky:2005bw,Dai:2014jja,Munoz:2014eqa,Cook:2015vqa}, where $\omega_{re}= -1/3$ corresponds to the end of inflation, and in order to satisfy the dominant energy condition of general relativity and maintain causality, $\omega_{re}$ must be less than 1 \cite{Munoz:2014eqa,Mishra:2021wkm,Chavanis:2014lra}. However, the EOS $\omega_{re}$ should vary with time during the reheating stage due to the non-equilibrium nonlinear dynamics of the field \cite{Podolsky:2005bw,Saha:2020bis}. Therefore, Ref. \cite{Saha:2020bis} discusses the time evolution equation of EOS during this stage and obtain a time-varying EOS equation, which alleviates the arbitrariness of defining EOS parameters during reheating. To this end, we will take the average $\omega_{re}$ during reheating for subsequent discussion based on the analysis of the evolution equation of EOS between the coherent oscillation and the radiation-dominated period in Ref. \cite{Saha:2020bis}, and give more details in Sec.{\uppercase\expandafter{\romannumeral4}}. B.

Furthermore, the number of e-folding $N_k$ from the end of inflation to the start of the radiation era is usually chosen to define the duration of reheating. The value of e-foldings $N_k$ is affected by the potential of inflation, the universe reheating instantaneously affects the upper limit of e-folding numbers, and the reheating temperature under the electroweak scale determines the lower limit. The value of $N_k$ can be between 46 and 70 to deal with the horizon problem \cite{Pareek:2021lxz}, and according to the analysis in Refs. \cite{Dodelson:2003vq,Liddle:2003as}, $N_k$ can even be 107 in some extreme cases.

The Natural Inflation(NI) model was first proposed in Ref.\cite{Freese:1990rb}. It has been a research hotspot in this field for several years because of its simple and clear formula, and it also produced the mass of pseudo-Goldstone bosons through non-perturbation effects. Moreover, the NI model has the shift symmetry, which can prevent the influence of radiative correction on potential \cite{Adams:1992bn}. Unfortunately, due to the limitation of the tensor-to-scalar ratio $r$, recent Planck+BICEP/Keck observations have ruled out the NI model \cite{Planck:2018jri,BICEP:2021xfz}. Since then, a large number of modified NI models have appeared \cite{Cheng:2021qmc,Cheng:2021nyo,Antoniadis:2018yfq,Nomura:2017ehb,Hong:2017ooe,Ferreira:2018nav,Simeon:2020lkd,Reyimuaji:2020goi,Salvio:2021lka,Zhang:2021ppy}, in this paper, we will study the three modified single-filed NI models and consider the constraints of CMB and reheating for three models.

The main contents of this article are as follows: In Sec. II, we will review the method of parameterization of reheating and derive the expressions of the reheating temperature $T_{re}$ and reheating duration $N_{re}$. In Sec. III, we derive $r$, $n_s$, and reheating parameters for three modified single-field NI models. In Sec. IV, we will explore the CMB and reheating constraints on those models and discuss their feasible intervals to satisfy the experimental conditions. In Sec. V, is reserved for a summary.

\section{Reheating}
After inflation is over, the energy of the universe exists in the scalar field. At this point, the temperature of the universe drops, and nucleosynthesis is pushed beyond the trigger boundary. Reheating is a transitional stage after the end of the inflation, which can release the energy in the scalar field and heat the universe, thereby ensuring the smooth appearance of the radiation-dominated period. As mentioned before, the reheating phase can be parameterized as temperature $T_{re}$, effective state equation $\omega_{re}$ and duration e-folding number $N_{re}$.

Next, we will give the derivations of $T_{re}$ and $N_{re}$ in detail from inflation models \cite{Easther:2011yq,Dai:2014jja,Mielczarek:2010ag}. According to the energy density evolution equation in the inflation universe, we can get $\rho\propto\alpha^{-3(1+\omega)}$, and
\begin{eqnarray}
\frac{\rho_{end}}{\rho_{re}}&=&(\frac{a_{end}}{a_{re}})^{-3(1+\omega_{re})},
\label{eq1}
\end{eqnarray}
where ``$end$'' and ``$re$'' represent the end of inflation and reheating, respectively. From Eq.\ref{eq1}, the e-folding number of reheating can be expressed as
\begin{eqnarray}
N_{re}=\frac{1}{3(1+\omega_{re})}\ln(\frac{\rho_{end}}{\rho_{re}}),
\end{eqnarray}
furthermore, $\omega_{re}=-\frac{1}{3}$ corresponds to the end of inflation, and one can get $\rho_{end}=\frac{3}{2}V_{end}$. After reheating, the universe will enter a period of radiation dominance and the energy density has the relationship with the reheating temperature $\rho_{re}=\frac{1}{30} \pi^{2} g_{re} T^{4}_{re}$. Where $g_{re}$ is dominant for the number of relativistic species at the end of reheating, and we use $g_{re}\approx100$ for the following discussion in this article. Therefore, the duration $N_{re}$ can be further expressed as a function of $T_{re}$
\begin{eqnarray}
N_{re}&=&\frac{1}{3(1+\omega_{re})} \ln(\frac{45V_{end}}{\pi^{2} g_{re} T^{4}_{re}}).
\label{eq4}
\end{eqnarray}
Considering the variation of the number of helical states in the radiant gas as a function of temperature \cite{Cook:2015vqa}, the relationship between the reheating temperature $T_{re}$ and today's temperature $T_{0}$ is obtained as
\begin{eqnarray}
T_{re}=T_{0}(\frac{a_{0}}{a_{re}})(\frac{43}{11g_{re}})^{\frac{1}{3}}=T_{0}(\frac{a_{0}}{a_{eq}})e^{N_{RD}}(\frac{43}{11g_{re}})^{\frac{1}{3}},
\label{eq5}
\end{eqnarray}
where the subscripts ``$eq$'' and ``$RD$'' represent the matter-dominated period and radiation-dominated epoch, respectively. And $e^{N_{RD}}=\frac{a_{eq}}{a_{re}}$ with the length in e-folds of radiation dominance $N_{RD}$. The time to cross the Hubble radius during inflation is represented by pivot scale $k=a_{k}H_{k}$ including a Hubble parameter during the inflation $H_k$, thus we can rewrite the ratio $a_0/a_{eq}$ into
\begin{eqnarray}
\frac{a_{0}}{a_{eq}}&=&\frac{a_{0}H_{k}}{k}e^{-N_{k}}e^{-N_{re}}e^{-N_{RD}},
\label{eq6}
\end{eqnarray}
where $e^{N_k}=a_{end}/a_{k}$ and $e^{N_{re}}=a_{re}/a_{end}$, and $``k"$ denotes the value of Fourier mode $k$ when it leaves the Hubble radius during inflation. Then Eq.\ref{eq5} can be rewritten as
\begin{eqnarray}
T_{re}&=&(\frac{43}{11g_{re}})^{\frac{1}{3}}(\frac{a_{0}T_{0}}{k})H_{k}e^{-N_{k}}e^{-N_{re}}.
\label{eq7}
\end{eqnarray}
Two special cases need to be considered, e.g., $\omega_{re}= \frac{1}{3}$ and $\omega_{re}\neq \frac{1}{3}$. First, assuming $\omega_{re}\neq \frac{1}{3}$ and putting Eq.\ref{eq7} into Eq.\ref{eq4}, one can get
\begin{eqnarray}
N_{re}&=&\frac{4}{1-3\omega_{re}}[-\frac{1}{4}\ln(\frac{45}{\pi^{2}g_{re}})-\ln(\frac{V^{\frac{1}{4}}_{end}}{H_{k}})\nonumber\\&& -\frac{1}{3}\ln(\frac{11g_{re}}{43})-\ln(\frac{k}{a_{0}T_{0}})-N_{k}].
\label{eq8}
\end{eqnarray}
If we choose the Planck pivot $0.05Mpc^{-1}$, the Eq.\ref{eq8} simplifies to
\begin{eqnarray}
N_{re}&=&\frac{4}{1-3\omega_{re}}[61.6-\ln(\frac{V^{\frac{1}{4}}_{end}}{H_{k}})-N_{k}].
\label{eq9}
\end{eqnarray}
Likewise, Eq.\ref{eq7} can also be abbreviated as
\begin{eqnarray}
T_{re}&=&[(\frac{43}{11g_{re}})^{\frac{1}{3}} \frac{a_{0}T_{0}}{k} H_{k} e^{-N_{k}} (\frac{45V_{end}}{\pi^{2}g_{re}})^{-\frac{1}{3(1+\omega_{re})}}]^{\frac{3(1+\omega_{re})}{3\omega_{re}-1}}.
\label{eq10}
\end{eqnarray}
In the second case, e.g., $\omega_{re}=\frac{1}{3}$, Eq.\ref{eq4} becomes
\begin{eqnarray}
0&=&\frac{1}{4}\ln(\frac{30}{\pi^{2}g_{re}})+\frac{1}{4}\ln(\frac{3}{2})+\ln(\frac{V^{\frac{1}{4}}_{end}}{H_k})+\frac{1}{3}\ln(\frac{11g_{re}}{43})\nonumber\\&&
+\ln(\frac{k}{a_{0}T_{0}})+N_{k},
\label{eq811}
\end{eqnarray}
and if one chooses $g_{re}=100$, then the above formula can be simplified to
\begin{eqnarray}
61.55&=&\ln(\frac{V^{\frac{1}{4}}_{end}}{H_k})+N_k.
\label{eq81}
\end{eqnarray}
Since $\omega_{re}=\frac{1}{3}$ corresponds to the start of the radiation-dominated period, it's impossible to obtain the expressions for $N_{re}$ and $T_{re}$, but we can obtain the constraints on $n_s$ for a particular model.

\section{INFLATON POTENTIALS}
The theoretical motivation for the NI model is clear and simple in form, but it is contradicted by observational data with more than $95\%$ confidence, especially with the recently published experimental data of Planck+BICEP/Keck \cite{Planck:2018jri,BICEP:2021xfz}. Based on this, many studies on modification of the NI model have been reported \cite{Cheng:2021qmc,Cheng:2021nyo,Antoniadis:2018yfq,Nomura:2017ehb,Hong:2017ooe,Ferreira:2018nav,Simeon:2020lkd,Reyimuaji:2020goi,Salvio:2021lka,Zhang:2021ppy}, which is expected to match the experimental data. This paper focuses on three modified single-field NI models, we derive the tensor-to-scalar ratio $r$ and spectral index $n_s$ of the models, and investigate the reheating constraints on these models.

\subsection{Special generalized Natural Inflation}
According to the Generalized Natural Inflation \cite{Cheng:2021qmc},
\begin{eqnarray}
V(\phi)&=&\Lambda^{4}[\cos \frac{\phi}{f_{m}}+\varepsilon \cos \frac{\phi}{f_{m}}+ e^{\frac{1}{n_{1}!}\cos(\frac{\phi}{f_{m}}+\frac{\pi}{n_{1}!})^{n_{1}}}]^{n_{2}},
\label{eq131}
\end{eqnarray}
a special generalized natural inflation(SNI) model can be obtained with $n_1=\pm \infty$, $n_2=1$ and $\varepsilon=0$, where $\Lambda^{4}$ represents the energy density and $f_m$ is the decay constant.

The e-folds number $N_{k}$ is defined as
\begin{eqnarray}
N_{k}&=&\frac{1}{M^{2}_{p}}\int_{\phi_{end}}^{\phi_{k}}\frac{V}{V^{'}} d\phi,
\label{151}
\end{eqnarray}
where the subscript $``end"$ refers to the value of the inflation field at the end of inflation. According to the definition of slow-roll parameters $\epsilon=\frac{M^{2}_{p}}{2}(\frac{V^{'}}{V})^{2}$ and $\eta=M^{2}_{p}\frac{V^{''}}{V}$, choosing $\epsilon=1$ as the end of inflation, we can derive
\begin{eqnarray}
\phi_{end}&=&\frac{1}{f_{m}}\arccos(\sqrt{\frac{M^{2}_{p}}{2f_{m}^{2}+M^{2}_{p}}}).
\label{eq122}
\end{eqnarray}
By using the spectral index $n_{s} = 1-6\epsilon+ 2\eta$ and $r=16\epsilon$ at $\phi=\phi_{k}$, this will lead to
\begin{eqnarray}
n_{s}&=&\frac{f_{m}^{2}-M^{2}_{p}[3\tan^{2}(\frac{\phi_{k}}{f_{m}})+2]}{f_{m}^{2}},
\label{eq123}
\end{eqnarray}
and
\begin{eqnarray}
r&=&-\frac{8[f_{m}^{2}(n_{s}-1)+2M^{2}_{p}]}{3f_{m}^{2}}.
\label{eq124}
\end{eqnarray}
According to $H_{k}=\pi M_{p}\sqrt{8A_{s}\epsilon}$ and $V\approx3H^{2}_{k} M^{2}_{p}$, the Hubble parameter $H_k$ can be directly deduced as
\begin{eqnarray}
H_{k}&=&\frac{2\sqrt{3}\pi M_{p}\sqrt{-A_{s}[f^{2}_{m}(n_{s}-1)+2M^{2}_{p}]}}{3f_{m}},
\label{22}
\end{eqnarray}
and the potential of the end of the inflation $V_{end}$ becomes
\begin{eqnarray}
V_{end}&=&4\pi^{2}A_{s} M^{4}_{p}(-2\frac{M^{2}_{p}}{f^{2}_{m}}-n_{s}+1)\frac{\cos(\frac{\phi_{end}}{f_{m}})}{\cos(\frac{\phi_{k}}{f_{m}})},
\label{23}
\end{eqnarray}
where the expression of $\phi_k(n_s)$ can be obtained by inversely solving Eq.\ref{eq123}, the scalar amplitude $A_{s}\approx2.196\times 10^{-9}$ and $ M_{p}=2.4\times10^{18} {\rm GeV}$ \cite{Planck:2015sxf}.

\subsection{Extended Natural Inflation model}
The second model, we can call it the ``Extended'' Natural Inflation(ENI) model. Where the inflaton is the pseudo-Nambo-Goldstone boson and the shift symmetry preserves the flatness of the potential, it can be written as \cite{Munoz:2014eqa}
\begin{eqnarray}
V(\phi)&=&\frac{2\Lambda^{4}}{2^{m}}(1+\cos \frac{\phi}{f_{e}})^{m},
\label{eq13}
\end{eqnarray}
where $f_{e}$ is the decay constant for the ENI model, and the application scope of the model can be expanded by changing the value of the parameter $m$. Following the method in the previous section,  the $r$, $n_s$, $H_k$ and $V_{end}$ of the ENI model can be obtained as follows:
\begin{eqnarray}
r&=&-8m\frac{f_{e}^{2}(n_{s}-1)+mM^{2}_{p}}{f_{e}^{2}(m+1)},
\label{21}
\end{eqnarray}
\begin{eqnarray}
n_{s}&=&1-\frac{mM^{2}_{p}\sec^{2}(\frac{\phi_{k}}{2f_{e}})[m\cos(\frac{\phi_{k}}{f_{e}})-m-2]}{2f_{e}^{2}},
\end{eqnarray}

\begin{eqnarray}
H_{k}&=&2\pi M_{p} \sqrt{-A_{s} \frac{m[f_{e}^{2}(n_{s}-1)+m M^{2}_{p}]}{f_{e}^{2}(m+1)}},
\end{eqnarray}
and
\begin{eqnarray}
V_{end}&=&-12\pi^{2}A_{s}m^{2}M^{4}_{p}\frac{[m^{2}M^{2}_{p}-f_{e}^{2}(n_{s}-1)]}{f_{e}^{2}(m+1)^{2}}\nonumber\\&&
\times \frac{[f_{e}^{2}(n_{s}-1)+m M^{2}_{p}]}{(2f_{e}^{2}+m^{2}M^{2}_{p})}.
\label{34}
\end{eqnarray}

\subsection{Natural Inflation inspired model}
The third model is the Natural Inflation inspired(NII) model, which is able to make the spontaneous symmetry breaking scale less than 1 \cite{German:2021jer} and its potential can be expressed as
\begin{eqnarray}
V(\phi)&=&V_{0}[1-\sin^{2}(\frac{\phi}{f_{n}})],
\end{eqnarray}
where $f_{n}$ is the decay parameter for the NII model and one can do the same steps as the SNI model, it is obtained
\begin{eqnarray}
n_s&=&1-2\frac{M^{2}_{p}}{f_{n}^{2}}[\cos(\frac{2\phi_{k}}{f_{n}})-3]\sec^{2}(\frac{\phi_{k}}{f_{n}}),
\end{eqnarray}

\begin{eqnarray}
r=16\epsilon=-\frac{2[f_{n}^{2}(n_{s}-1)+4M^{2}_{p}]^{2}}{f_{n}^{4}(n_{s}-1)},
\label{32}
\end{eqnarray}

\begin{eqnarray}
H_{k}&=&\sqrt{2}\pi M_{p} \sqrt{A_{s}(-\frac{4M^{2}_{p}}{f_{n}^{2}}-n_{s}+1)},
\label{33}
\end{eqnarray}
and
\begin{eqnarray}
V_{end}=-\frac{3 \pi^{2}A_{s} M^{4}_{p}[4M^{2}_{p}-f_{n}^{2}(n_{s}-1)][4M^{2}_{p}+f_{n}^{2}(n_{s}-1)]}{2f_{n}^{2}(f_{n}^{2}+2M^{2}_{p})}.
\end{eqnarray}

\section{Results}
\subsection{CMB constraints}
\begin{figure}[htb]
\centering
\includegraphics[width=0.5\textwidth]{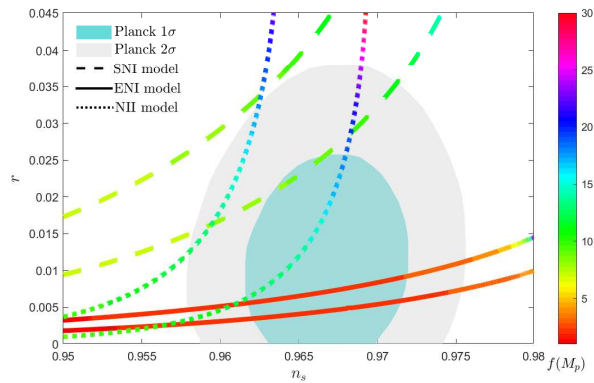}
\caption{The relationship between $r$ and $n_s$ for three single-field NI models under different $N_k\in[55,65]$. The dashed line, solid and dotted lines represent the relationship between $n_{s}-r$ under the SNI model, ENI model, and NII model, respectively. Among the ENI model, we choose $m=0.1$ for subsequent discussion. The colored gradient graph corresponds to the value of decay constant varying from $1 M_{p}$ to $30 M_{p}$. In addition, the light blue and light grey broadband correspond to the latest combination $1\sigma$ and $2\sigma$ experiment errors of BAO, BICEP/KECK, and Planck data, respectively \cite{BICEP:2021xfz}.}
\label{fig0}
\end{figure}

Fig.~\ref{fig0} shows the relationship between $n_{s}$ and $r$ under different $N_k$ for SNI, ENI, and NII. For each model, we take two different e-folding numbers, which $N_k=55$ and $N_k=65$ correspond to up and down lines, respectively. The Gradient graph indicates the value of decay constant varying from $1 M_{p}$ to $30 M_{p}$. The light blue and light grey shaded broadband stand for the latest experiments with $1\sigma$ and $2\sigma$ experiment errors of BAO, BICEP/KECK, and Planck data, respectively \cite{BICEP:2021xfz}.

In Fig.~\ref{fig0}, the dashed line represents the feasible parameter space of $n_s$ and $r$ of the SNI model and $N_{k}\in[55,65]$. In the SNI model, the $r$ value matches the experimental data within 2$\sigma$ error when $N_k \geq 55$, and the value of $r$ matches the experimental data within 1$\sigma$ error when $N_k\geq65$. Furthermore, with the constrained by CMB, the decay constants are in the range of $8 M_p -10 M_p$. The solid line represents the feasible parameter space of the ENI model and we will choose $m=0.1$ for the following discussion. In the ENI model, the parameter spaces of $r$ and $n_s$ are consistent with the latest experimental data with 1$\sigma$ error when $N_{k}$ varies in $[55-65]$ and the decay constant is constrained to be less than $2 M_p$. The $r$ and $n_s$ of the NII model dependence curves under different e-folds numbers are shown as dotted lines of Fig.~\ref{fig0}. When $N_{k}\geq55$, the obtained results are within the error of the $2\sigma$ experimental data. In addition, when $N_k>60$, $r$ can falls within the experimental boundary of 1$\sigma$ Planck data. Moreover, the CMB constrained decay constant is in the range of $12 M_p -20 M_p$.

Fig.\ref{fig0} graphically shows that the NII model is more general than the SNI model since the feasible space of the SNI model on the $(n_s, r)$ plane is almost covered by NII under the constraints of the CMB. In addition, there's no overlapping part in the space of $(n_s, r)$ between the ENI model and the other two models, which indicates the ENI model is excluded by the other two models, and we need more accurate experiments to confirm it.

\begin{figure}[htb]
\centering
\includegraphics[width=0.5\textwidth]{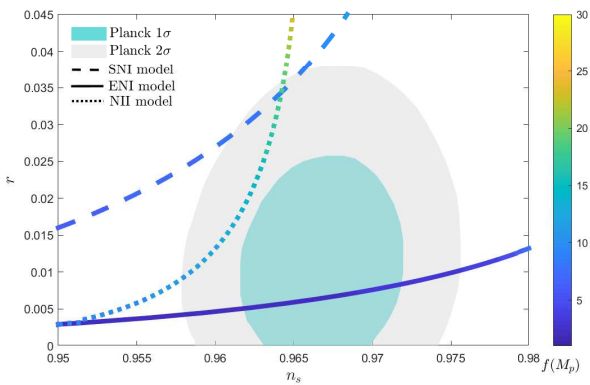}
\caption{The relationship between $r$ and $n_s$ for three single-field NI models under the special case $\omega_{re}=\frac{1}{3}$ and the constraints of the amplitude of scalar fluctuations. The dashed line, solid and dotted lines represent the relationship between $n_{s}-r$ under the SNI model, ENI model, and NII model, respectively. The colored gradient graph corresponds to the value of decay constant varying from $1 M_{p}$ to $30 M_{p}$. In addition, the light blue and light grey broadband correspond to the latest combination $1\sigma$ and $2\sigma$ experiment errors of BAO, BICEP/KECK, and Planck data, respectively \cite{BICEP:2021xfz}.}
\label{fig1}
\end{figure}

When the values of $H_{k}=\pi M_{p}\sqrt{8A_{s}\epsilon}$, $N_k$ and $V_{end}$ for the three models are brought into
\begin{eqnarray}
\ln(\frac{V^{\frac{1}{4}}_{end}}{H_k})+N_k-61.55,
\label{eq881}
\end{eqnarray}
then we can get the $(n_s,r)$ for $\omega_{re}=\frac{1}{3}$ and constraints of the amplitude of scalar fluctuations $A_s$, as shown in the Fig.\ref{fig1}. Where the dashed line, solid and dotted lines represent the relationship between $n_{s}- r$ under the SNI model, ENI model, and NII model, respectively. Likewise, the colored gradient graph indicates the decay constant varies from $1 M_{p}$ to $30 M_{p}$. The above conditions impose strong constraints on the parameter space. For the SNI model, only when the decay constant $8 M_p<f_{m}<10 M_p$,  $n_s$ and $r$ would be within the 2$\sigma$ experiment boundary.

It can be seen from Fig.\ref{fig1} that for any chosen value of $f_{e}$ under the ENI model, the $r$ can satisfy the constraints given by the latest experiments, however, $n_s$ is very sensitive to the change of parameter $f_{e}$. Under the constraints of the amplitude of scalar fluctuations, $n_s$ of the ENI model would be within experimental error when $f_{e} \rightarrow 2 M_p$ and varies in a small range \cite{BICEP:2021xfz}. In addition, when $f_{e}>4 M_{p}$, the $r$ and $n_s$ of the ENI model tend to be stable and change in a small range.

Under the constraint of amplitude of scalar fluctuations, the variation of $r$ and $n_s$ with $f_{n}$ is shown in Fig.\ref{fig1}. Where the the dotted line represents the change curve of the NII model. As Fig.\ref{fig1} shows, those above conditions impose strong constraints on the parameter space. Only when $f_{n}\in[14 M_{p},20 M_{p}]$, $n_s$ and $r$ can be within the range of the latest 2$\sigma$ Planck experiment \cite{BICEP:2021xfz}, far from the experimental range of 1$\sigma$.

\subsection{Reheating constraints}
The value of EOS can be fixed by the inflation model parameter $p$ when the scalar field oscillates near the minimum potential at the end of inflation, i.e. the EOS of a homogeneous condensate oscillating in potential with a minimum of the form $V(\phi) \propto \phi^{p}$ can be parameterized as $\omega=(p-2)/(p+2)$, and this allows us to naturally derive the value of EOS at this stage \cite{Podolsky:2005bw,Lozanov:2017hjm,Turner:1983he,Mishra:2021wkm,Saha:2020bis}. However, at the end of the coherent oscillation phase, fragmentation leads to inhomogeneities, which in turn alter the EOS during the phase of backreaction \cite{Saha:2020bis,Micha:2002ey,Micha:2004bv,Amin:2010dc,Amin:2014eta,Maity:2018qhi}. The effects of fragmentation on the evolution equation of EOS at this stage can usually be obtained through lattice simulation, see \cite{Amin:2010dc,Maity:2018qhi,Lozanov:2016hid,Lozanov:2017hjm} for more details. Furthermore, the Ref. \cite{Saha:2020bis} studies the EOS in the reheating phase after coherent oscillation in detail, and gives a more precise EOS for this stage.

Based on this, as an attempt, we also analyze the minimum potential behavior of the three models listed in this paper. However, we found that there is a constant correction term in the expanded form of the three models, and the effect of this correction may require lattice calculations to estimate, which is very challenging work. Fortunately, the value of EOS has a manageable impact on subsequent research in this paper. Therefore, assuming that the constant coefficient correction is small enough and we can infer the $V(\phi) \propto \phi^{2}$ near the minimum potential for the three models. Then, the EOS at different stages can be obtained naturally based on Ref. \cite{Saha:2020bis}, i.e. $\omega_{re}=-\frac{1}{3}$, $\omega_{re}=0$, $\omega_{re}=\frac{1}{5}$ and $\omega_{re}=1$, where $\omega_{re}=0$ stands for the coherent oscillation stage and $\omega_{re}=1/5$ for the reheating stage.

\begin{figure*}[tb]
\centering
\subfigure{
\includegraphics[width=0.45\textwidth]{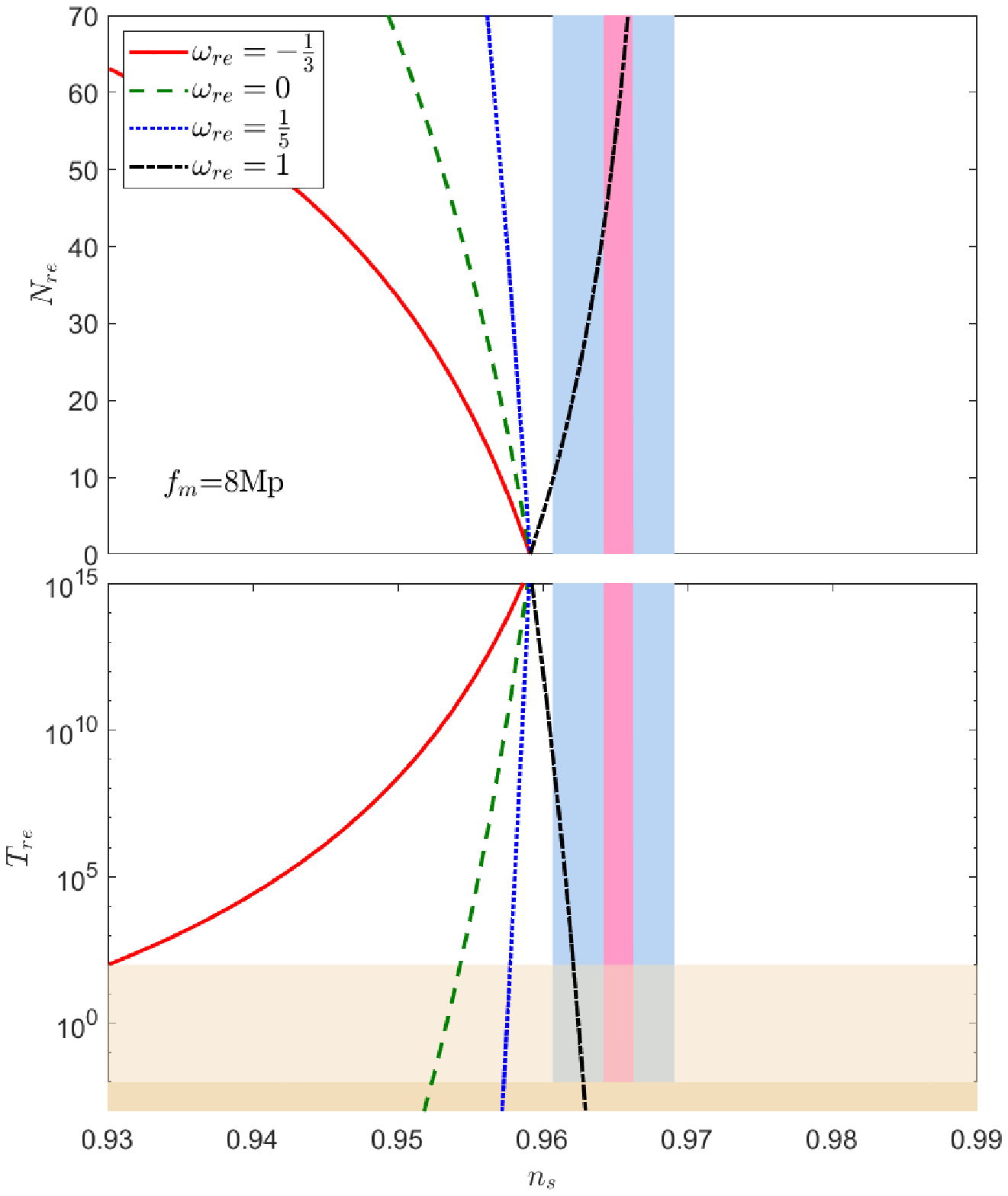}
}
\quad
\subfigure{
\includegraphics[width=0.45\textwidth]{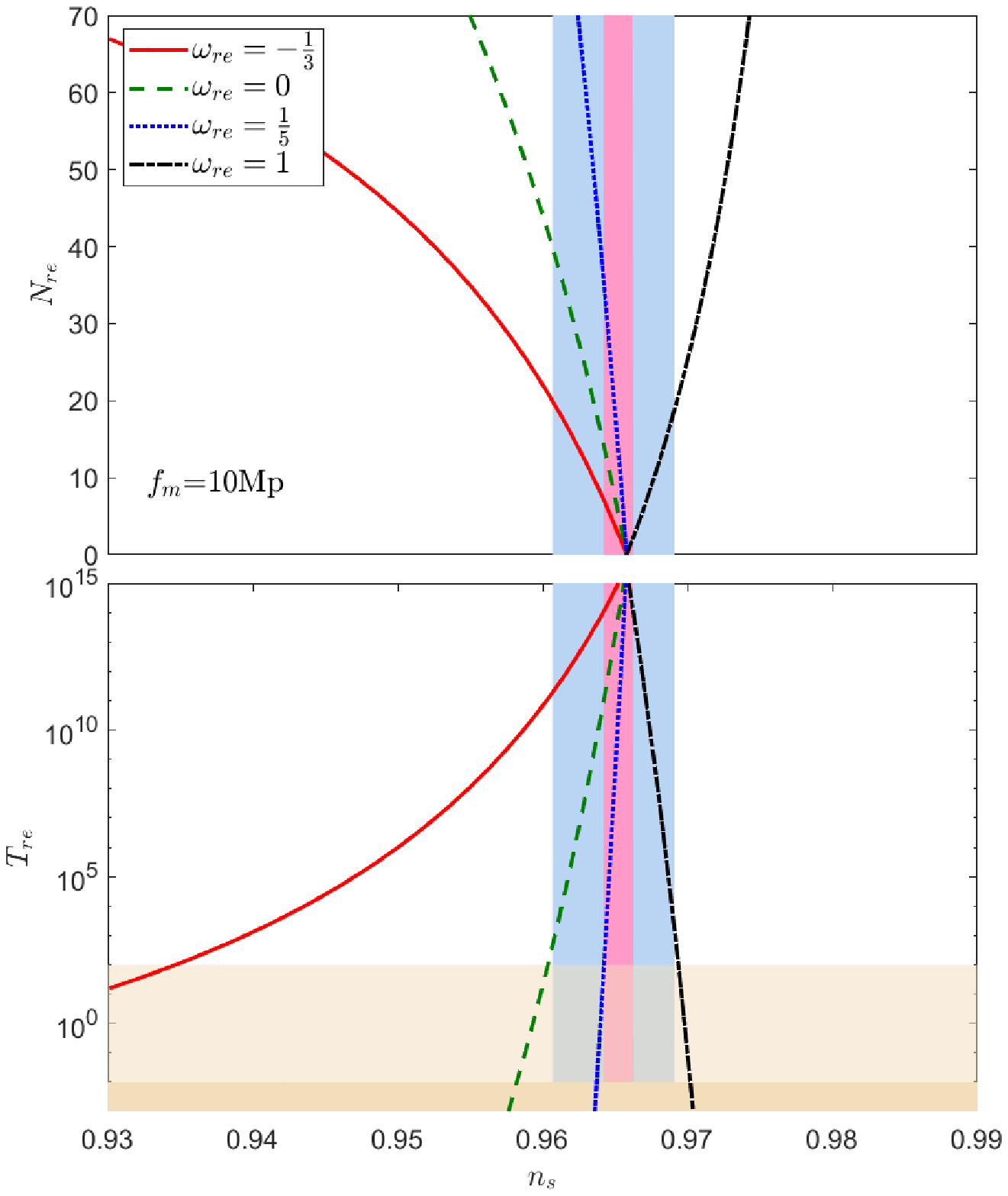}
}
\subfigure{
\includegraphics[width=0.45\textwidth]{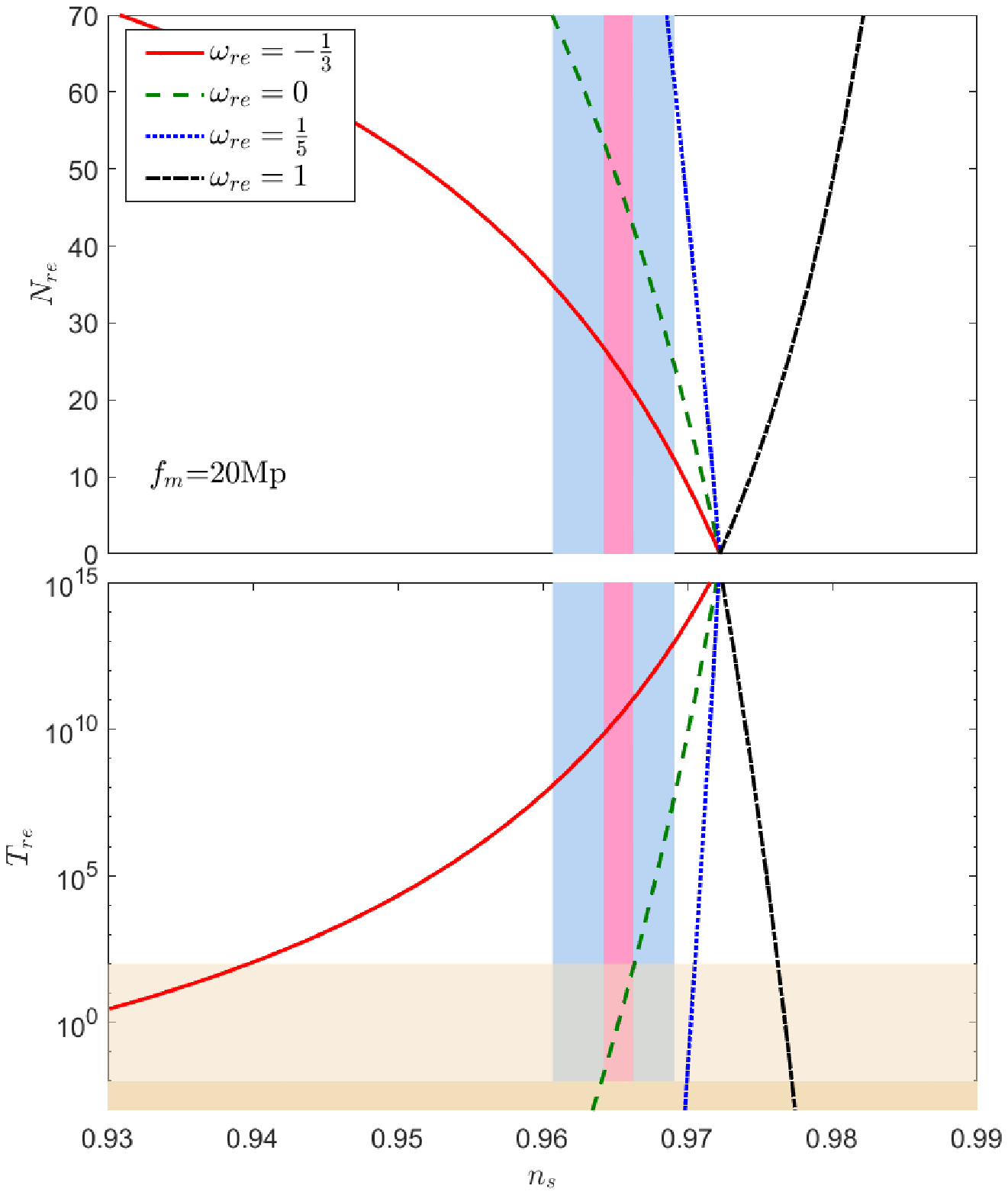}
}
\quad
\subfigure{
\includegraphics[width=0.45\textwidth]{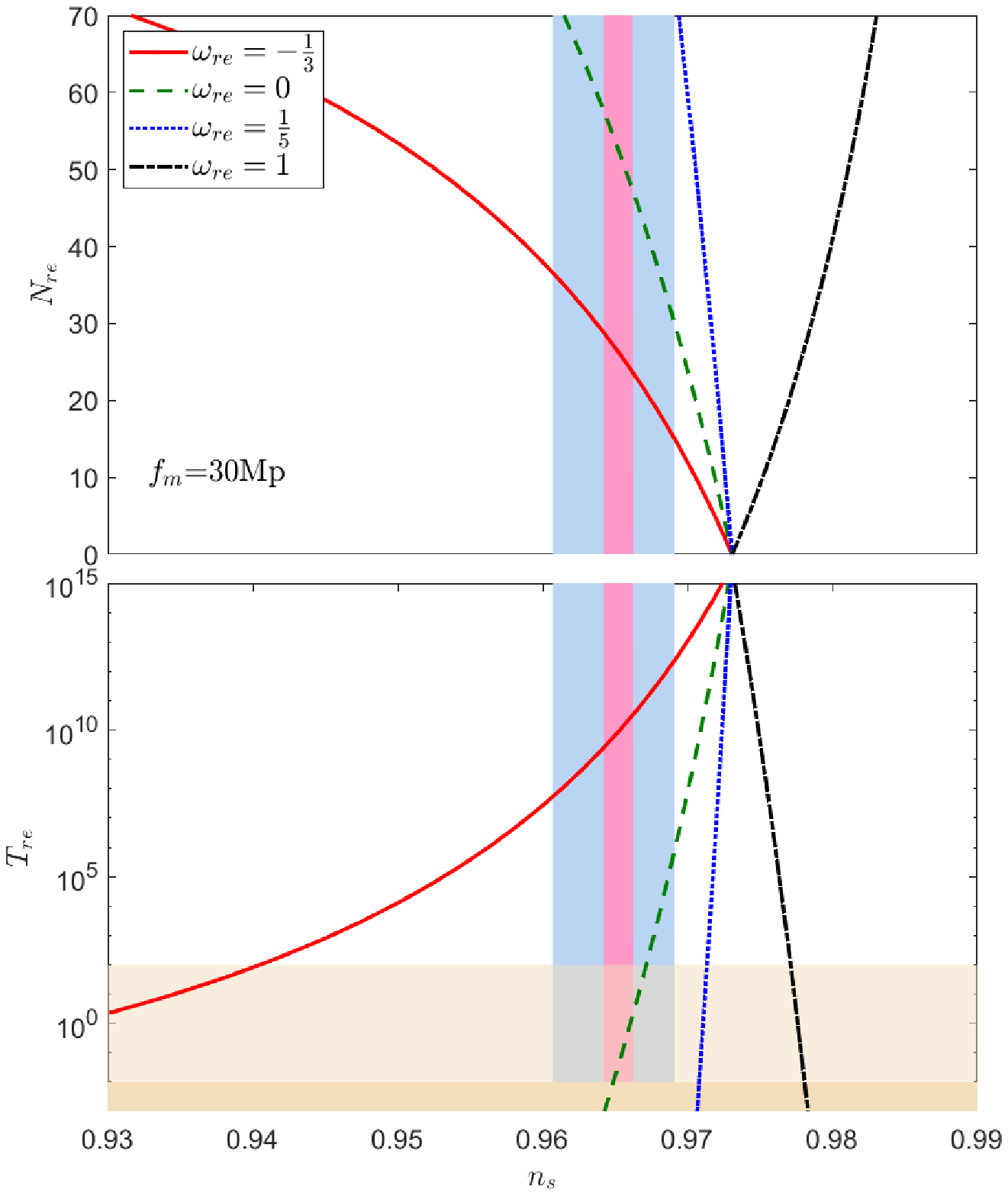}
}
\caption{$T_{re}$ and $N_{re}$ as a function of $n_s$ for different $f_{m}$ and $\omega_{re}$ in the SNI model. The red solid, the gree dashed, the blue dotted and the black dotted and dashed line corresponds to $\omega_{re}=-\frac{1}{3}$, $\omega_{re}=0$, $\omega_{re}=\frac{1}{5}$ and $\omega_{re}=1$, respectively. The blue region corresponds to the $1\sigma$ boundary of Planck $n_s$ and the red area corresponds to the $1\sigma$ boundary of the further CMB experiment with sensitivity $\pm10^{-3}$ \cite{EuclidTheoryWorkingGroup:2012gxx,PRISM:2013ybg}. The khaki area corresponds to temperatures of 10 MeV from BBN, and Light khaki areas correspond to electroweak scales below 100 GeV.}
\label{fig50}
\end{figure*}

The behaviors of $N_{re}$ and $T_{re}$ as a function of $n_s$ under the SNI model as shown in Fig.\ref{fig50}, the blue region corresponds to the $1\sigma$ boundary on Planck's $n_s$ and the red one corresponds to the further experiment precision of $10^{-3}$. According to the constraints of CMB on the model parameter space, we choose four typical values of $f_{m}$ for subsequent discussion, i.e., $f_{m}=8 M_p$, $f_{m}=10 M_p$, $f_{m}=20 M_p$ and $f_{m}=30 M_p$. The $T_{re}$ converges around $10^{15}$ GeV, which may be required by the GUT-scale regeneration model \cite{Bourakadi:2021hbn}. The point where the four lines come together ($N_{re}=0$) is what be called the instantaneous reheating point \cite{Pareek:2021lxz}.

\begin{figure*}[tb]
\centering
{
\includegraphics[width=0.48\textwidth]{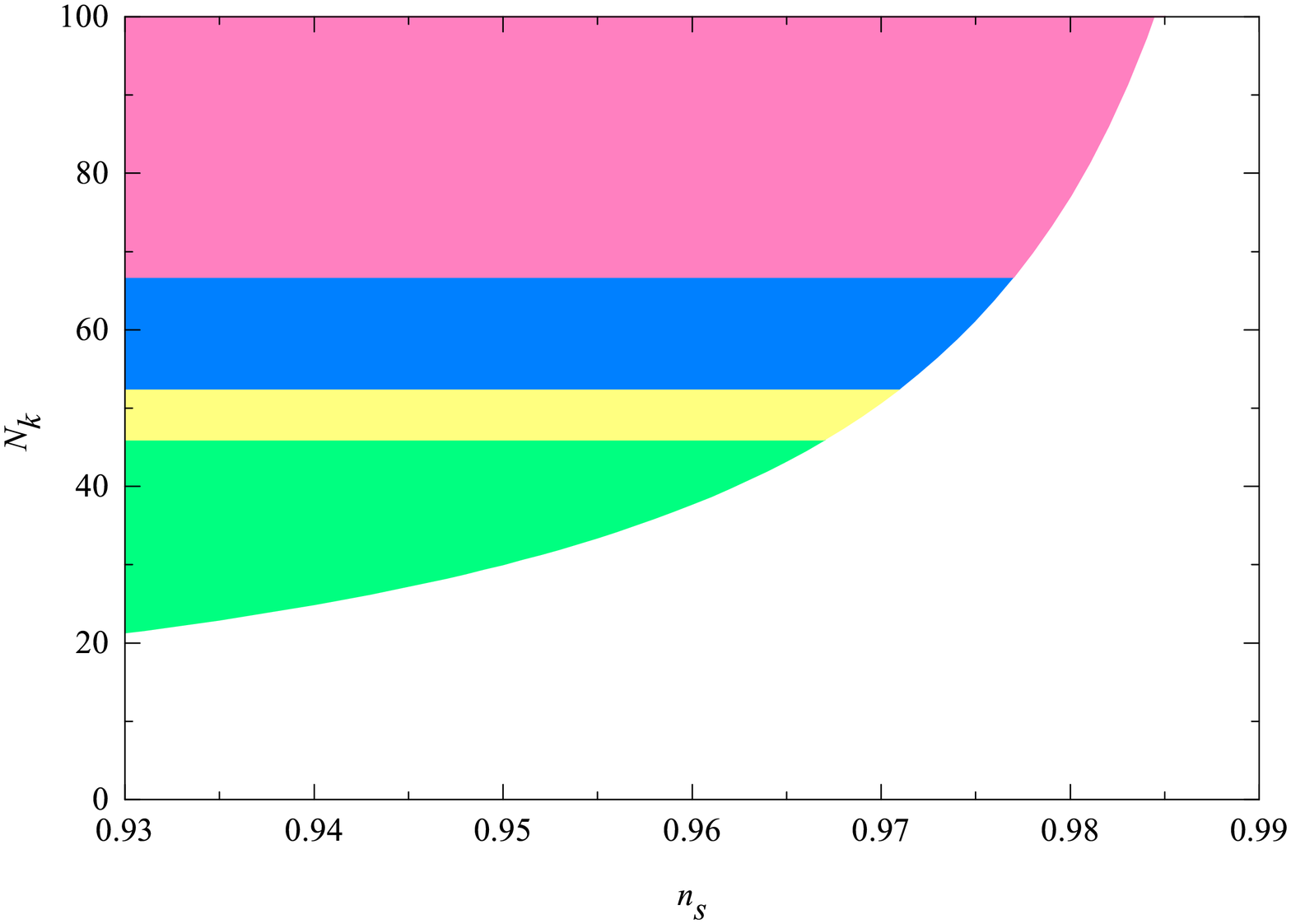}
}
\quad
{
\includegraphics[width=0.48\textwidth]{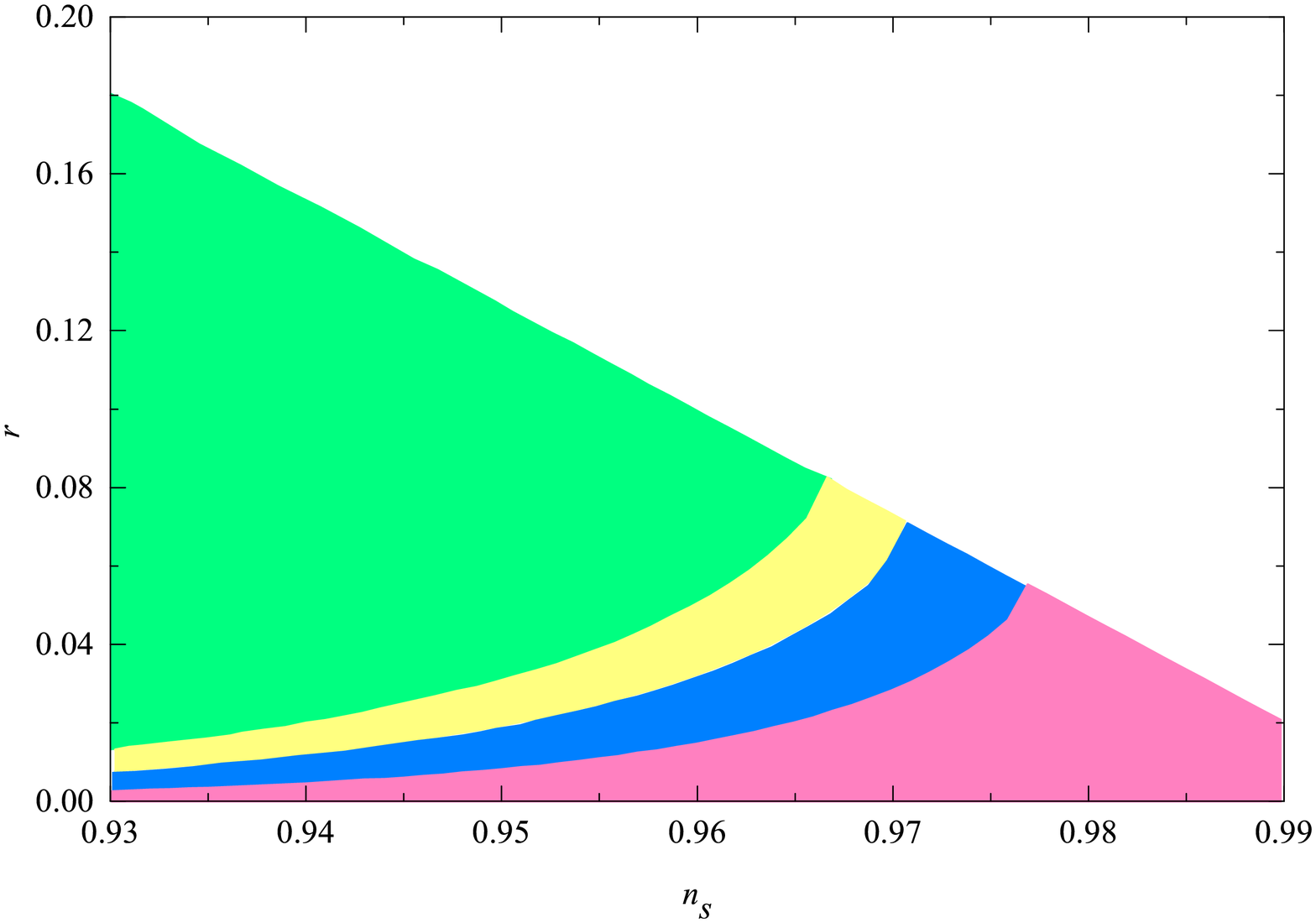}
}
\caption{The $N_{k}$ vs $n_{s}$ and $r$ vs $n_{s}$ under the SNI model, where $f_{m}<30 M_p$. The green area corresponds to $\omega_{re} \leq0$, the yellow area corresponds to $0 \leq \omega_{re} \leq \frac{1}{5}$, the blue area represents the range of $\frac{1}{5} \leq \omega_{re} \leq 1$, and $\omega_{re} \geq1$ corresponds to the dark pink range.}
\label{fig40}
\end{figure*}

The relationship between $N_{k}-n_{s}$ and $r-n_s$ is obvious in Fig.\ref{fig40}. The green area corresponds to $\omega_{re} \leq0$, the yellow area corresponds to $0 \leq \omega_{re} \leq \frac{1}{5}$, the blue area represents the range of $\frac{1}{5} \leq \omega_{re} \leq 1$, and $\omega_{re} \geq1$ corresponds to the dark pink range. Since the value of $f_{m}$ is proportional to $n_s$, the lower bound of the $N_{k}-n_{s}$ corresponds to larger $f_{m}$-values, conversely, the upper bound of the $r-n_s$ corresponds to larger $f_{m}$. One can also find that, both $r$ and $n_s$ are within the Planck-2018 constraint when $\omega_{re}>0$. If one sets $T_{re}=100$ GeV, thus the bounds of $n_s$, $N_k$ and $r$ all can be obtained from the constraints with different $f_{m}$ and $\omega_{re}$. From Table.\ref{tab30} one can get that $\omega_{re}$ is proportional to $n_s$ and $N_k$, and inversely proportional to $r$. Furthermore, under such restriction of $T_{re}$, the corresponding $\omega_{re}$ can be found only when $f_{m} < 10 M_p$, one can find the solution that both $n_s$ and $r$ are within the latest Planck-2018 data \cite{BICEP:2021xfz}.

\begin{widetext}
\makeatletter\def\@captype{table}\makeatother
\begin{center}
\setlength{\tabcolsep}{3mm}
\renewcommand\arraystretch{1}
\begin{tabular}{|c|c |c |c| c |c| c}
\hline
  ~$f_{m}( M_{p})$         &~ $\omega_{re}$~           & ~$n_s$~    &~$N_k$~      &~r~\\
\hline
~                     &$[-\frac{1}{3},0]$           &$[0.9300,0.9543] $ &$ [25.12,46.04]$   &$ [0.1033,0.0385]$  \\
8                    &$[0,\frac{1}{5}]$          &$[0.9543,0.9578] $ &$ [46.04,52.93] $    &$ [0.0385,0.0293]$ \\
~                    &$[\frac{1}{5},\frac{1}{3}]$          &$ [0.9578,0.9591]$ &$ [52.93,56.37]$    &$ [0.0293,0.0257]$  \\
~                    &$[\frac{1}{3},1]$          &$ [0.9591,0.9621]$ &$ [56.37,66.62]$    &$ [0.0257,0.0176]$  \\
\hline
~                     &$[-\frac{1}{3},0]$           &$ [0.9347,0.9604] $ &$ [25.16,46.13]$   &$ [0.1208,0.0524]$  \\
10                    &$[0,\frac{1}{5}]$            &$[0.9604,0.9642]$ &$ [46.13,53.04]$   &$[0.0524,0.0420]$  \\
~                     &$[\frac{1}{5},\frac{1}{3}]$          &$[0.9642,0.9658]$ &$ [53.04,56.49] $   &$ [0.0420,0.0379]$ \\
~                    &$[\frac{1}{3},1]$          &$ [0.9658,0.9694]$ &$ [56.49,66.80]$    &$ [0.0379,0.0283]$  \\
\hline
~                     &$[-\frac{1}{3},0]$          &$ [0.9397,0.9663] $ &$ [25.20,46.24] $   &$ [0.1474,0.0764] $ \\
20                    &$[0,\frac{1}{5}]$           &$ [0.9663,0.9705] $ &$ [46.24,53.19] $   &$ [0.0764,0.0653] $ \\
~                     &$[\frac{1}{5},\frac{1}{3}]$        &$ [0.9705,0.9722]$ &$[53.19,56.66] $     &$ [0.0653,0.0608] $  \\
~                    &$[\frac{1}{3},1]$          &$ [0.9722,0.9762]$ &$ [56.66,67.03]$    &$ [0.0608,0.0500]$  \\
\hline
~                     &$[-\frac{1}{3},0]$          &$ [0.9397,0.9672] $  &$ [25.21,46.26] $  &$ [0.1548,0.0816] $   \\
30                    &$[0,\frac{1}{5}]$            &$[0.9672,0.9714]$  &$ [46.26,53.22]$   &$[0.0816,0.0705] $ \\
~                     &$[\frac{1}{5},\frac{1}{3}]$           &$[0.9714,0.9731]$ &$ [53.22,56.69]$   &$ [0.0705,0.0659]$ \\
~                    &$[\frac{1}{3},1]$          &$ [0.9731,0.9771]$ &$ [56.69,67.07]$    &$ [0.0659,0.0551]$  \\
\hline
\end{tabular}
\caption{The values of $n_s$, $N_k$ and $r$ under the SNI model corresponds to different values of $f_{m}$ and $\omega_{re}$, where $T_{re} = 100 $ GeV.}
\label{tab30}
\end{center}
\end{widetext}

\begin{figure*}[tb]
\centering
\subfigure{
\includegraphics[width=0.45\textwidth]{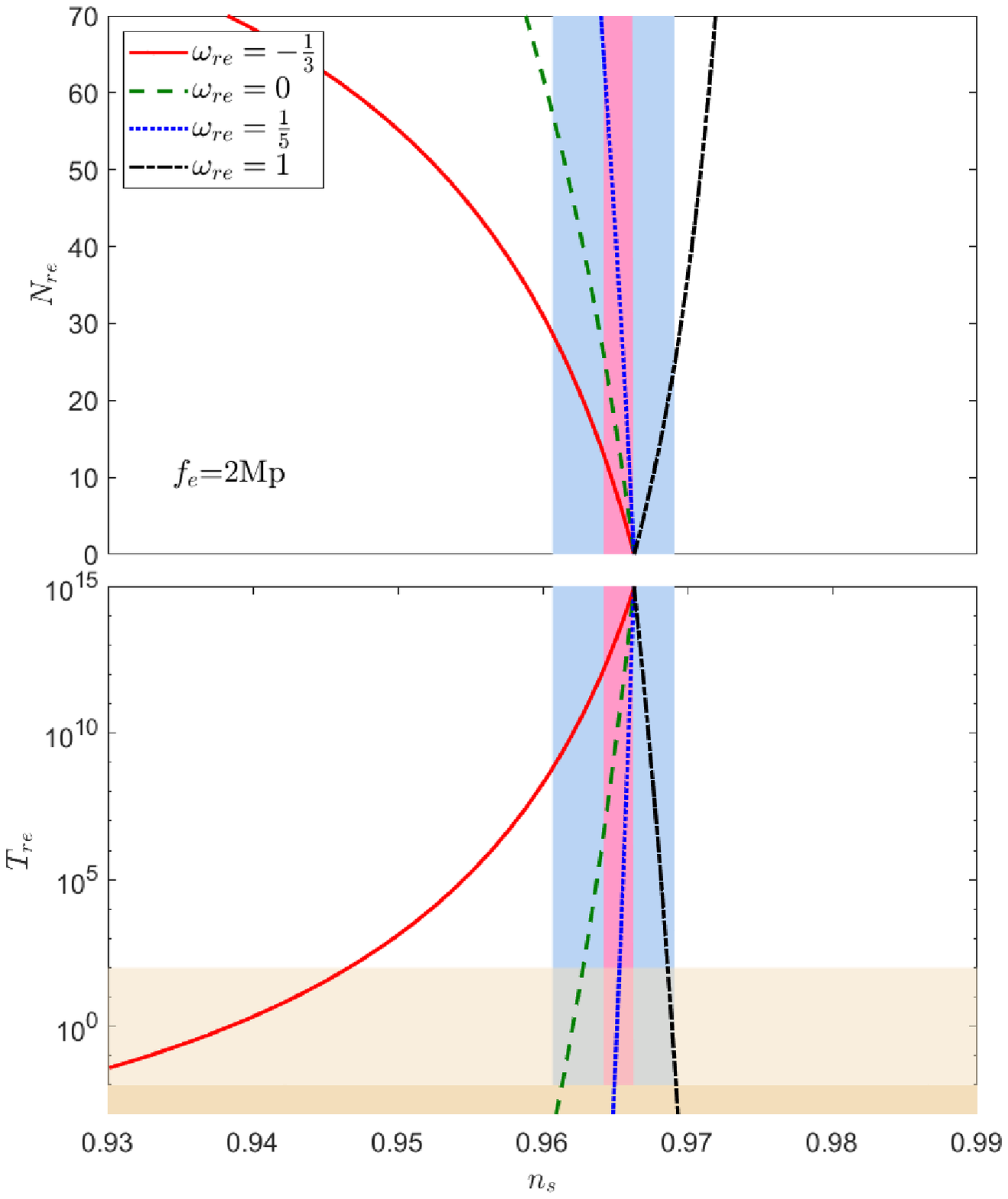}
}
\quad
\subfigure{
\includegraphics[width=0.45\textwidth]{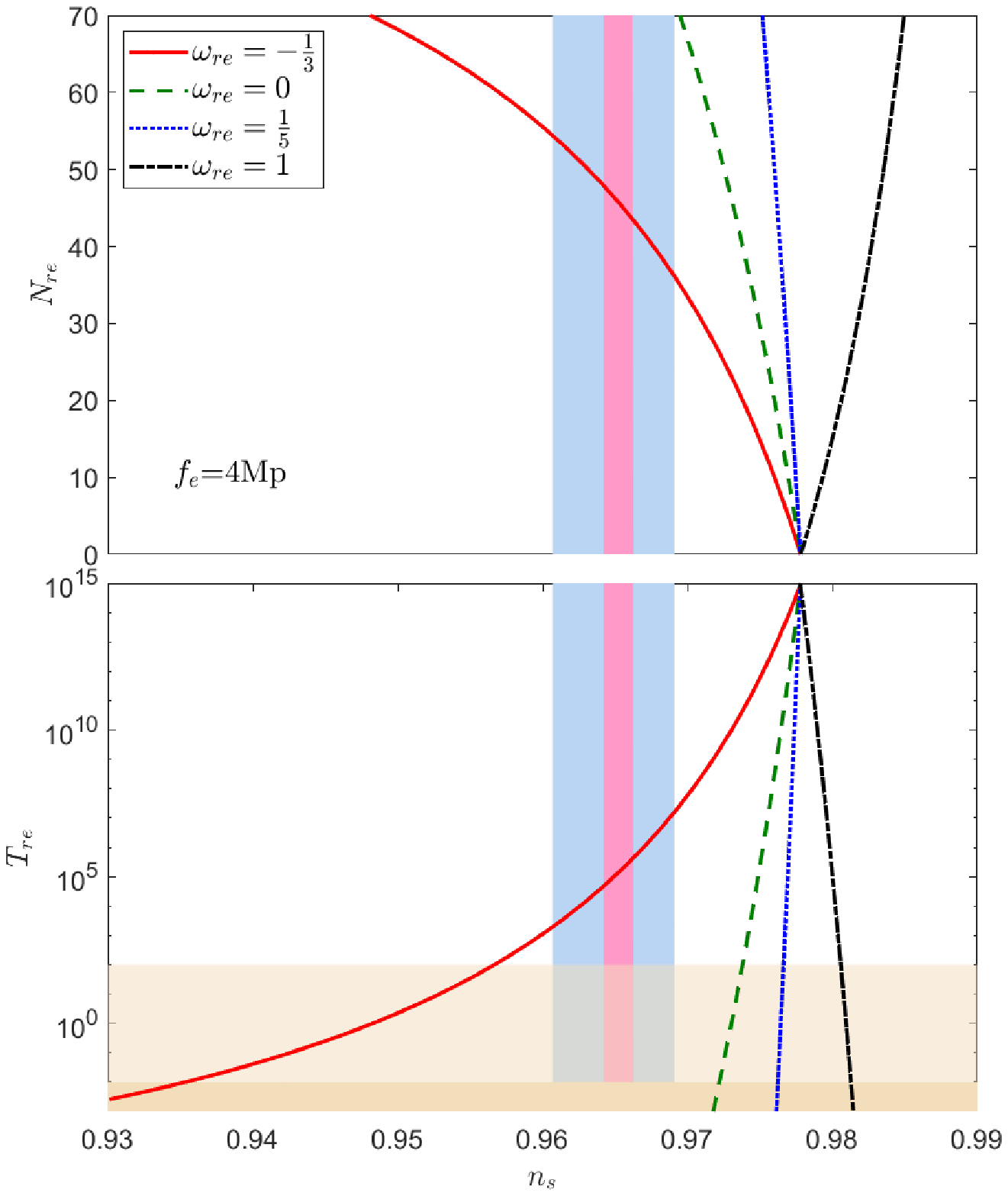}
}
\subfigure{
\includegraphics[width=0.45\textwidth]{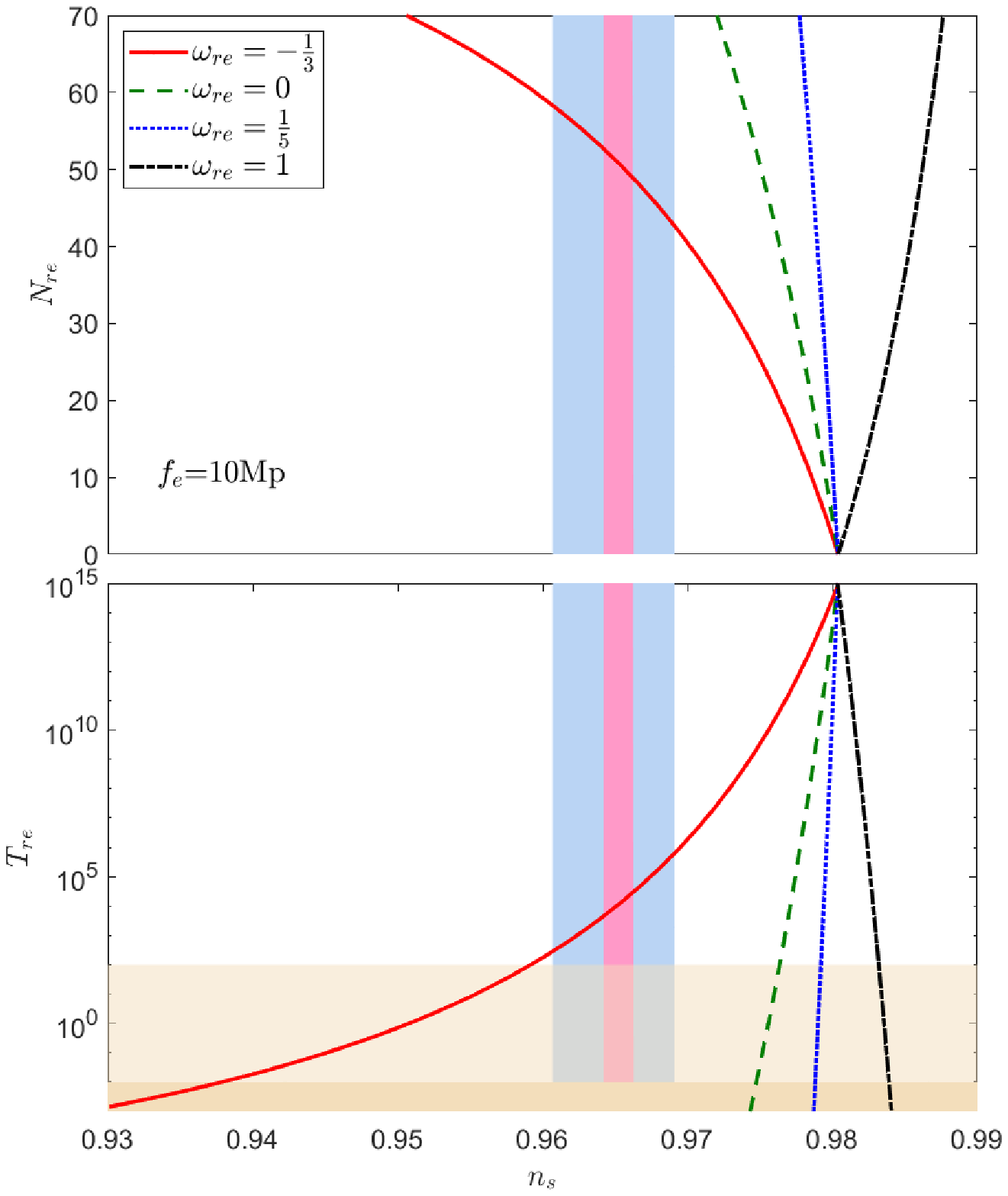}
}
\quad
\subfigure{
\includegraphics[width=0.45\textwidth]{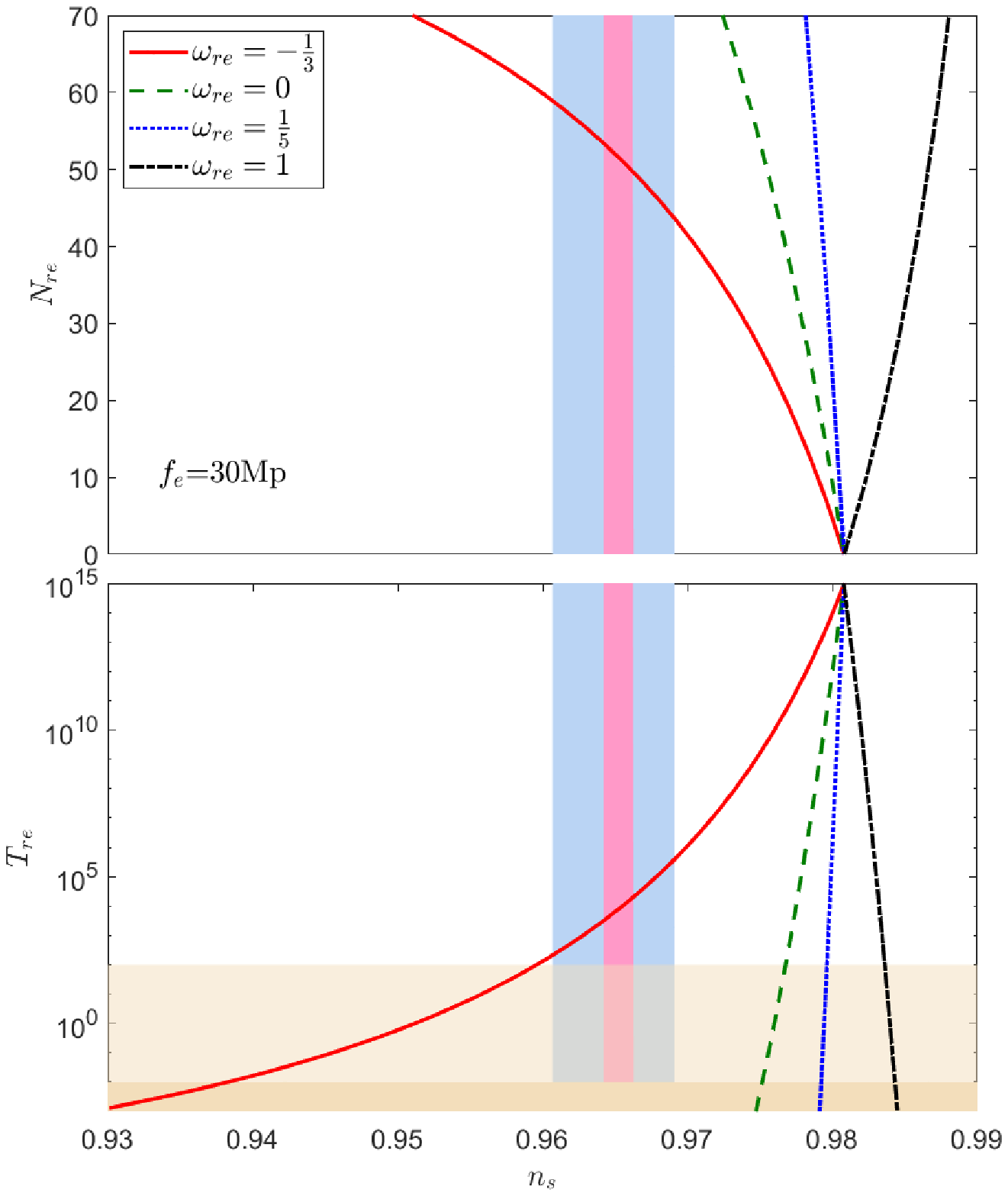}
}
\caption{The duration of reheating $N_{re}$ and the temperature $T_{re}$ as a function of $n_s$ are plotted for different parameters $f_{e}$ and $\omega_{re}$ of the ENI model. The legend of the shadow area refer to Fig.\ref{fig50}.}
\label{fig3}
\end{figure*}

\begin{figure*}[tb]
\centering
{
\includegraphics[width=0.48\textwidth]{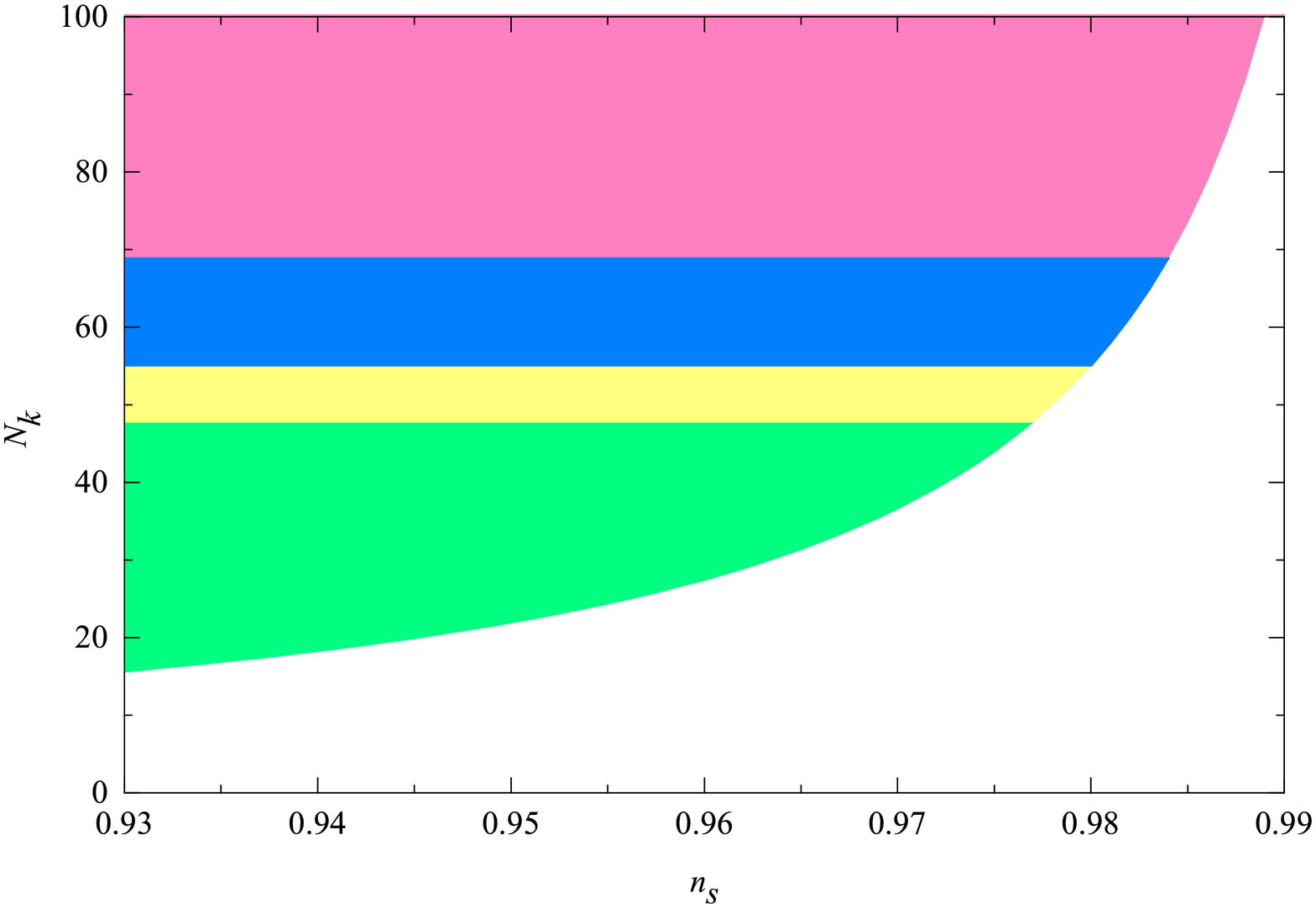}
}
\quad
{
\includegraphics[width=0.48\textwidth]{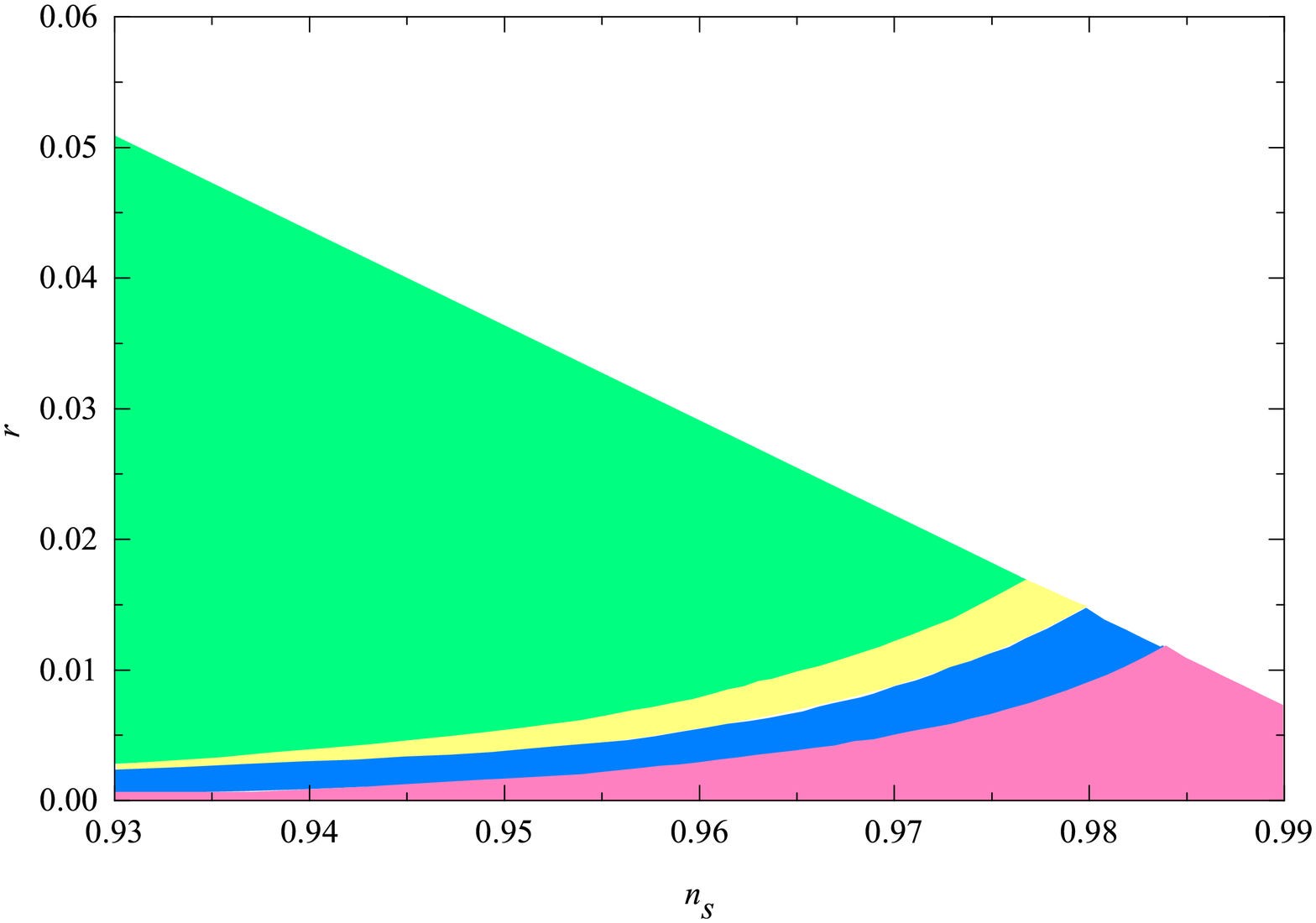}
}
\caption{$N_{k}$ vs $n_{s}$ and $r$ vs $n_{s}$ under the ENI model, where $f_{e}<30 M_p$. The legend of the shadow area refer to Fig.\ref{fig40}. }
\label{fig4}
\end{figure*}

Fig.\ref{fig3} shows $N_{re}$ and $T_{re}$ as a function of $n_s$ under the ENI model. By varying the decay constants $f_{e}$ in the feasible interval, e.g., $2 M_p$, $4 M_p$, $10 M_p$ and $30 M_p$, one can find that the ENI model is sensitive to the choice of parameter when $f_{e}<4 M_p$, i.e., $f_{e}>2 M_p$, the central value is rapidly away from the experimental error range and tends to be stable after $f_{e}>4 M_p$. Under four different $f_{e}$, the value of $n_s$, $r$ and $N_{k}$ for different $\omega_{re}$ can be found in Table.\ref{tab3}. when $f_{e} \geq 4 M_p$ and $0 \leq \omega_{re} \leq1$, $n_s$ is totally excluded by the Planck-2018. While, the value of $r$ satisfy the experimental constraints \cite{BICEP:2021xfz} for any chosen of $f_{e}< 30 M_{p}$ and $\omega_{re}\in[-1/3,1]$. Fig.\ref{fig4} shows the relationship between the predictions of $N_k-n_s$ and $r-n_s$ for different $\omega_{re}$, and one can find that when $\omega_{re}>0$, the value of $r$ can within the experimental error.

\begin{widetext}
\makeatletter\def\@captype{table}\makeatother
\begin{center}
\setlength{\tabcolsep}{3mm}
\renewcommand\arraystretch{1}
\begin{tabular}{|c|c |c |c| c |c| c}
\hline
  ~$f_{e}( M_{p})$         &~ $\omega_{re}$~           & ~$n_s$~    &~$N_k$~      &~r~\\
\hline
~                     &$[-\frac{1}{3},0]$           &$[0.9466,0.9628] $ &$ [27.05,47.14]$   &$ [0.0206,0.0089]$  \\
2                    &$[0,\frac{1}{5}]$          &$[0.9628,0.9653] $ &$ [47.14,53.74] $    &$ [0.0089,0.0070]$ \\
~                    &$[\frac{1}{5},\frac{1}{3}]$          &$ [0.9653,0.9663]$ &$ [53.74,57.03]$    &$ [0.0070,0.0063]$  \\
~                    &$[\frac{1}{3},1]$          &$ [0.9663,0.9686]$ &$ [57.03,66.86]$    &$ [0.0063,0.0046]$  \\
\hline
~                     &$[-\frac{1}{3},0]$           &$ [0.9567,0.9738] $ &$ [27.17,47.35]$   &$ [0.0270,0.0145]$  \\
4                    &$[0,\frac{1}{5}]$            &$[0.9738,0.9766]$ &$ [47.35,53.99]$   &$[0.0145,0.0124]$  \\
~                     &$[\frac{1}{5},\frac{1}{3}]$          &$[0.9766,0.9778]$ &$ [53.99,57.30] $   &$ [0.0124,0.0116]$ \\
~                    &$[\frac{1}{3},1]$          &$ [0.9778,0.9806]$ &$ [57.30,67.19]$    &$ [0.0116,0.0096]$  \\
\hline
~                     &$[-\frac{1}{3},0]$          &$ [0.9592,0.9764] $ &$ [27.20,47.41] $   &$ [0.0290,0.0165] $ \\
10                    &$[0,\frac{1}{5}]$           &$ [0.9764,0.9792] $ &$ [47.41,54.06] $   &$ [0.0165,0.0144] $ \\
~                     &$[\frac{1}{5},\frac{1}{3}]$        &$ [0.9792,0.9804]$ &$[54.06,57.37] $     &$ [0.0144,0.0135] $  \\
~                    &$[\frac{1}{3},1]$          &$ [0.9804,0.9832]$ &$ [57.37,67.28]$    &$ [0.0135,0.0115]$  \\
\hline
~                     &$[-\frac{1}{3},0]$          &$ [0.9596,0.9768] $  &$ [27.21,47.42] $  &$ [0.0293,0.0168] $   \\
30                    &$[0,\frac{1}{5}]$            &$[0.9768,0.9796]$  &$ [47.42,54.07]$   &$[0.0168,0.0147] $ \\
~                     &$[\frac{1}{5},\frac{1}{3}]$           &$[0.9796,0.9808]$ &$ [54.07,57.39]$   &$ [0.0147,0.0139]$ \\
~                    &$[\frac{1}{3},1]$          &$ [0.9808,0.9836]$ &$ [57.39,67.29]$    &$ [0.0139,0.0118]$  \\
\hline
\end{tabular}
\caption{The values of $n_s$, $N_k$ and $r$ under the ENI model corresponding to different $f_{e}$ and $\omega_{re}$, where $T_{re} = 100 $ GeV.}
\label{tab3}
\end{center}
\end{widetext}

\begin{figure*}[tb]
\centering
\subfigure{
\includegraphics[width=0.45\textwidth]{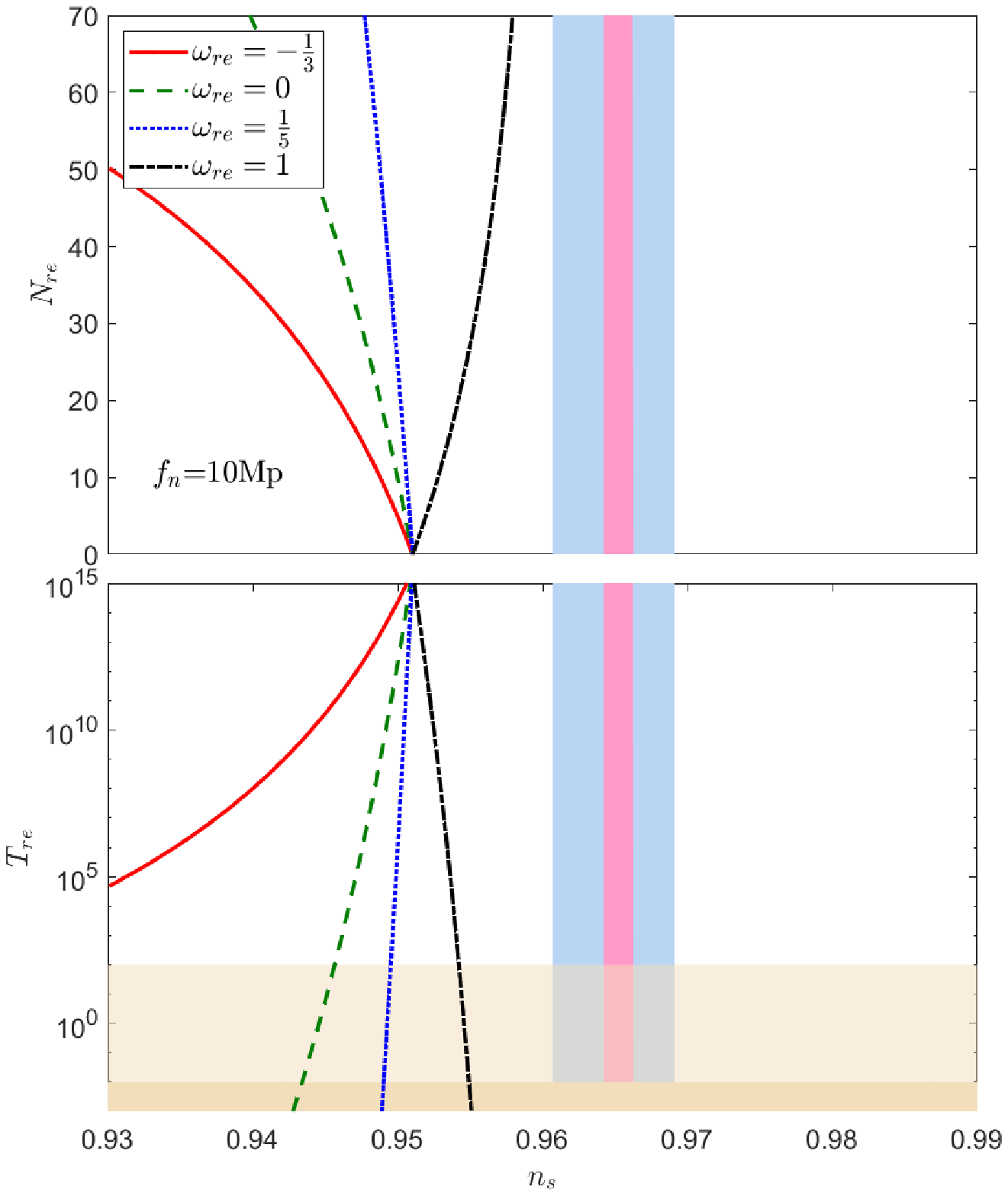}
}
\quad
\subfigure{
\includegraphics[width=0.45\textwidth]{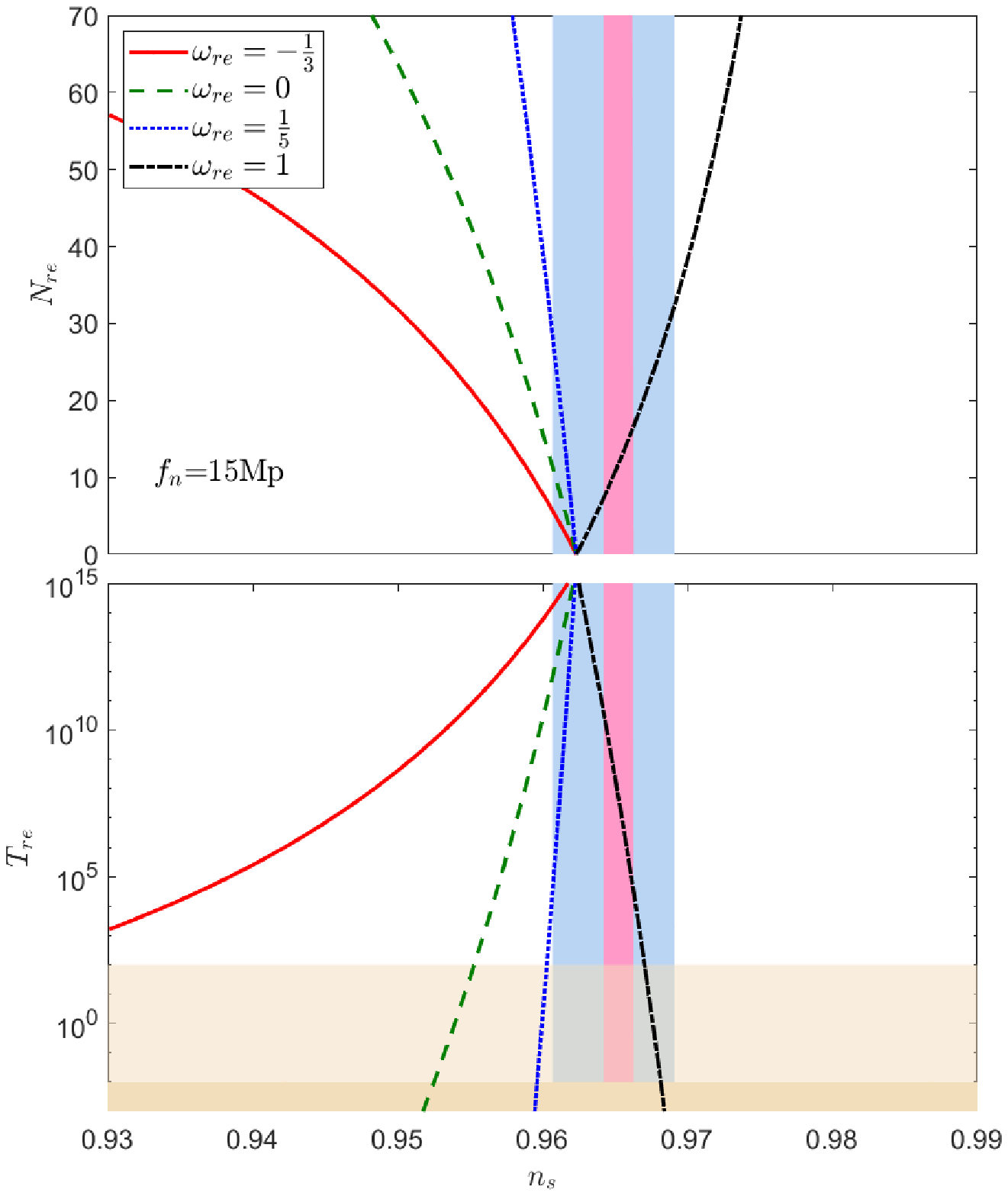}
}
\subfigure{
\includegraphics[width=0.45\textwidth]{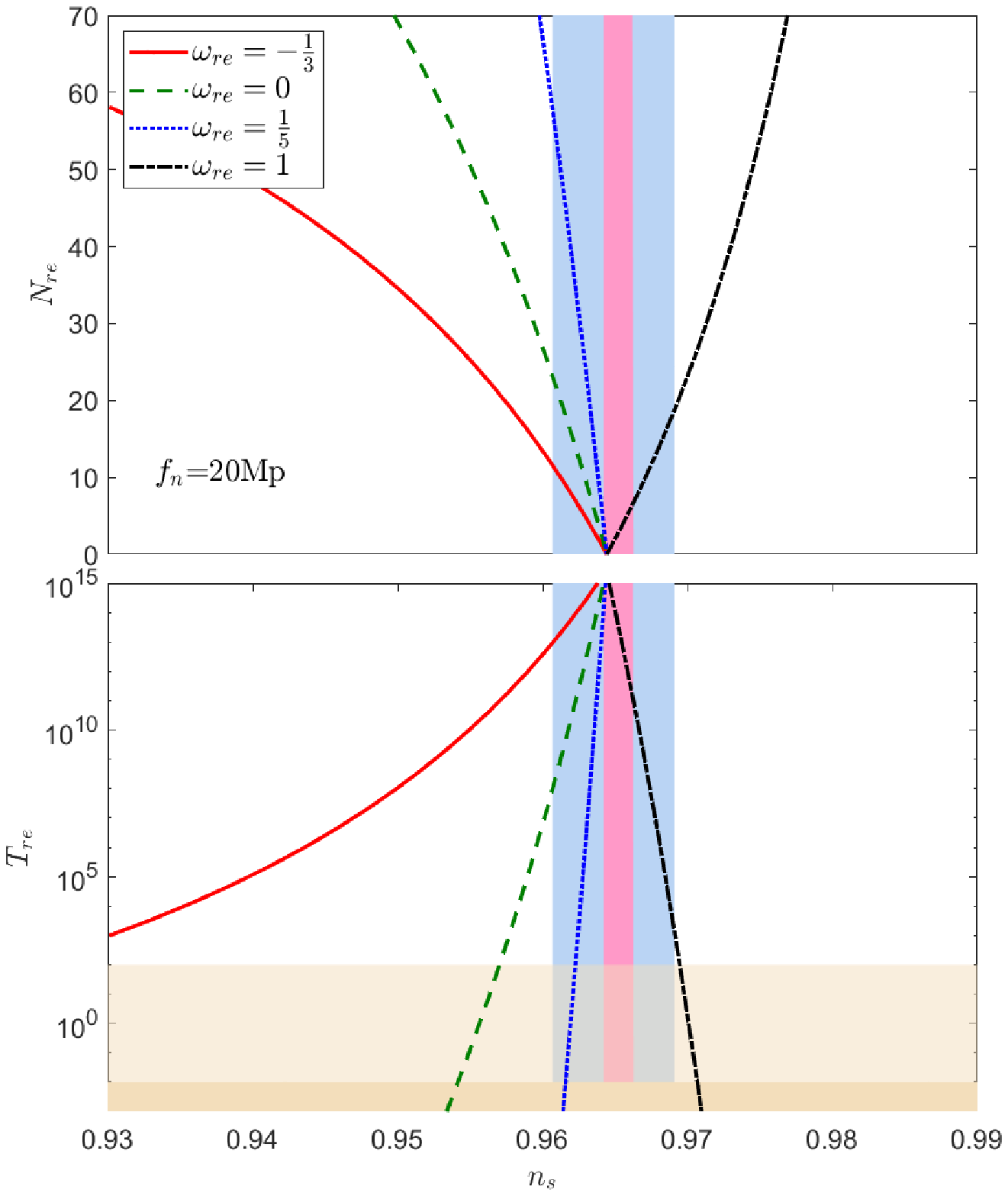}
}
\quad
\subfigure{
\includegraphics[width=0.45\textwidth]{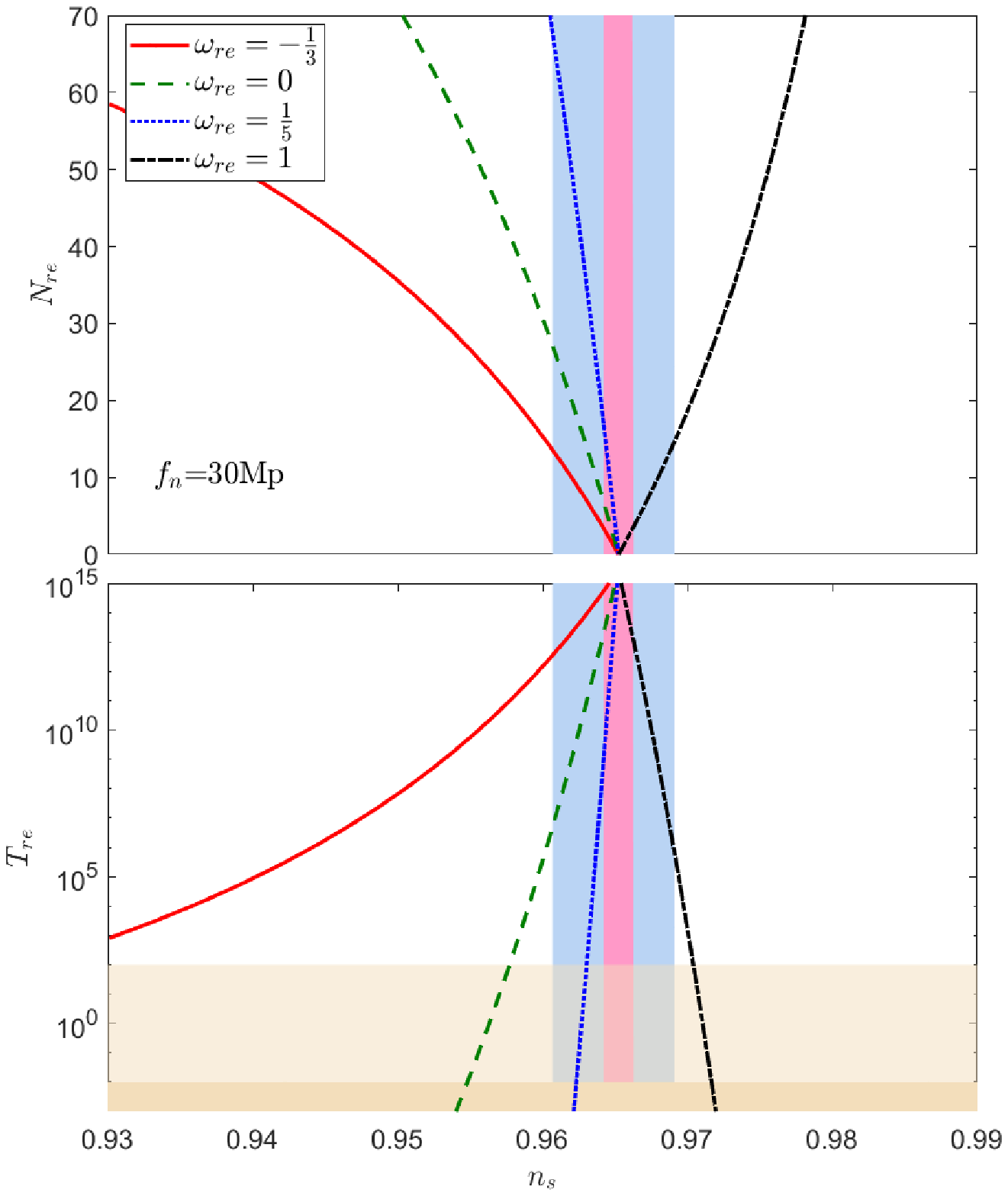}
}
\caption{$T_{re}$ and $N_{re}$ as a function of $n_s$ for different $f_{n}$ and $\omega_{re}$ in the NII model. The legend of the shadow area refer to Fig.\ref{fig50}.}
\label{fig5}
\end{figure*}

\begin{figure*}[tb]
\centering
{
\includegraphics[width=0.48\textwidth]{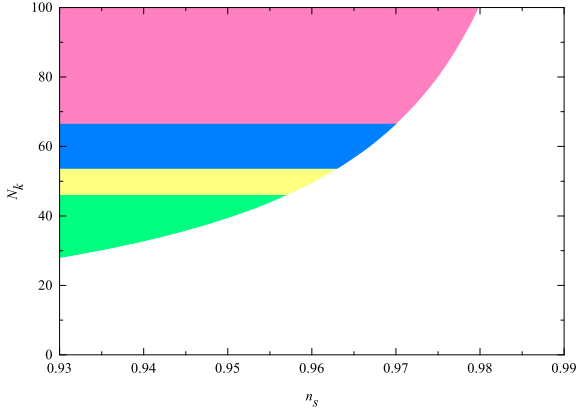}
}
\quad
{
\includegraphics[width=0.48\textwidth]{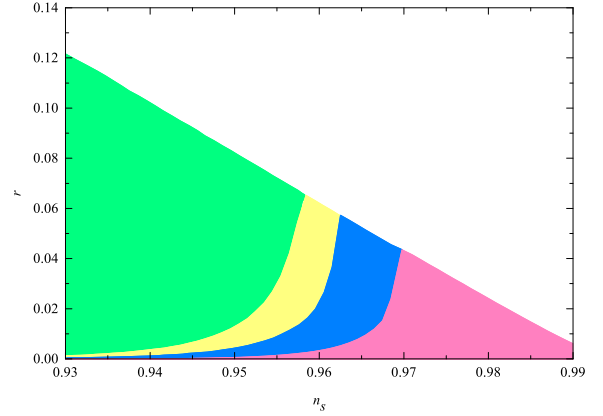}
}
\caption{$N_{k}$ vs $n_{s}$ and $r$ vs $n_{s}$ under NII model, where $f_{n}<30 M_p$. The legend of the shadow area refer to Fig.\ref{fig40}.}
\label{fig6}
\end{figure*}

The change of the temperature $T_{re}$ and the duration of reheating $N_{re}$ with $n_s$ under the NII model is shown in Fig.\ref{fig5}. Four typical $f_{n}$ have been chosen, i.e., $f_{n}=10 M_p$, $f_{n}=15 M_p$, $f_{n}=20 M_p$,and $f_{n}=30 M_p$, and one can find when $f_{n}>20 M_p$, the predictions of $N_{re}$ and $T_{re}$ tend to be stable. Table.\ref{tab4}. further show the variation of $n_s$ vs $r$ and $N_k$ vs $r$ under different effective equation of state $\omega_{re}$. It can be seen that when $f_{n} = 10 M_p$, the value of $r$ can fall within the experimentally allowable boundaries when $0<\omega_{re}<1$. However, even if a larger $\omega_{re}$ is chosen, the value of $n_s$ is still smaller than the experimental value. Fortunately, when $15 M_{p} \leq f_{n} \leq 20 M_p$, both $n_s$ and $r$ can within Planck-2018 error range under the constraints of $\frac{1}{5}\leq \omega_{re}\leq 1$ \cite{BICEP:2021xfz}. When $f_{n} > 30 M_p$, the values of $r$ is outside the experimental error range under any choice of $\omega_{re}$. In Fig.\ref{fig6}, it is graphically shows the changes of $N_{k}$ vs $n_{s}$ and $r$ vs $n_s$ under the constraints of $\omega_{re}$, especially when $0\leq\omega_{re} \leq \frac{1}{5}$, one can find the $n_s$ and $r$ lie within the contour constrained by Planck-2018.

\begin{widetext}
\makeatletter\def\@captype{table}\makeatother
\begin{center}
\setlength{\tabcolsep}{3mm}
\renewcommand\arraystretch{1}
\begin{tabular}{|c|c |c |c| c |c| c}
\hline
  ~$f_{n}( M_{p})$         &~ $\omega_{re}$~           & ~$n_s$~    &~$N_k$~      &~r~\\
\hline
~                     &$[-\frac{1}{3},0]$           &$[0.9171,0.9457] $ &$ [25.81,46.60]$   &$ [0.0444,0.0076]$  \\
10                    &$[0,\frac{1}{5}]$          &$[0.9457,0.9495] $ &$ [46.60,53.42] $    &$ [0.0076,0.0043]$ \\
~                    &$[\frac{1}{5},\frac{1}{3}]$          &$ [0.9495,0.9510]$ &$ [53.42,56.82]$    &$ [0.0043,0.0033]$  \\
~                    &$[\frac{1}{3},1]$          &$ [0.9510,0.9542]$ &$ [56.82,66.96]$    &$ [0.0033,0.0015]$  \\
\hline
~                     &$[-\frac{1}{3},0]$           &$ [0.9230,0.9553] $ &$ [25.94,46.84]$   &$ [0.0912,0.0325]$  \\
15                    &$[0,\frac{1}{5}]$            &$[0.9553,0.9603]$ &$ [46.84,53.71]$   &$[0.0325,0.0242]$  \\
~                     &$[\frac{1}{5},\frac{1}{3}]$          &$[0.9603,0.9623]$ &$ [53.71,57.13] $   &$ [0.0242,0.0210]$ \\
~                    &$[\frac{1}{3},1]$          &$ [0.9623,0.9670]$ &$ [57.13,67.37]$    &$ [0.0210,0.0140]$  \\
\hline
~                     &$[-\frac{1}{3},0]$          &$ [0.9240,0.9570] $ &$ [25.98,46.92] $   &$ [0.1146,0.0506] $ \\
20                    &$[0,\frac{1}{5}]$           &$ [0.9570,0.9623] $ &$ [46.92,53.81] $   &$ [0.0506,0.0408] $ \\
~                     &$[\frac{1}{5},\frac{1}{3}]$        &$ [0.9623,0.9644]$ &$[53.81,57.24] $     &$ [0.0408,0.0368] $  \\
~                    &$[\frac{1}{3},1]$          &$ [0.9644,0.9695]$ &$ [57.24,67.51]$    &$ [0.0368,0.0276]$  \\
\hline
~                     &$[-\frac{1}{3},0]$          &$ [0.9245,0.9577] $  &$ [26.02,46.99] $  &$ [0.1338,0.0677] $   \\
30                    &$[0,\frac{1}{5}]$            &$[0.9577,0.9630]$  &$ [46.99,53.89]$   &$[0.0677,0.0572] $ \\
~                     &$[\frac{1}{5},\frac{1}{3}]$           &$[0.9630,0.9652]$ &$ [53.89,57.32]$   &$ [0.0572,0.0529]$ \\
~                    &$[\frac{1}{3},1]$          &$ [0.9652,0.9704]$ &$ [57.32,67.60]$    &$ [0.0529,0.0427]$  \\
\hline
\end{tabular}
\caption{The values of $n_s$, $N_k$ and $r$ under different $f_{n}$ and $\omega_{re}$ of the NII model, where $T_{re} = 100 $ GeV.}
\label{tab4}
\end{center}
\end{widetext}

\begin{figure}[tb]
\centering
\includegraphics[width=0.5\textwidth]{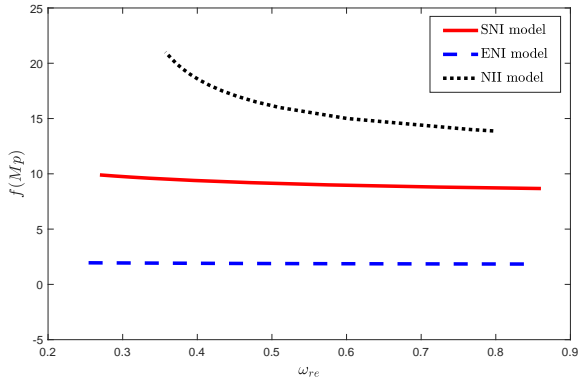}
\caption{Feasible parameter space comparison of SNI, ENI, and NII models under CMB and reheating constraints.}
\label{fig10}
\end{figure}

Combining the constraints of CMB and reheating, Fig.\ref{fig10} shows the value of the effective state equations $\omega_{re}$ of reheating for SNI, ENI, and NII models at different values of decay constants. Where the red solid line, the blue dashed line, and the black dotted line correspond to the SNI, ENI, and NII models, respectively. One can find that the decay constants of the three models are mutually exclusive. Furthermore, for the SNI model, the $n_{s}$, $r$ and $N_k$ all satisfy the constraints of the Planck-2018 when the eos $0.27<\omega_{re}<0.86$. For the ENI and NII models, in order to satisfy the constraints of the $n_{s}$, $r$ and $N_k$, $\omega_{re}$ should be within $0.24<\omega_{re}<0.84$ and $0.36<\omega_{re}<0.80$, respectively.

\section{Summary}
Reheating is an important period of inflation theory, which can release the energy stored in the scalar field at the end of inflation and increase the temperature of the universe. In this work, we study the evolution of the reheating after the end of inflation and investigate the constraints of the CMB and reheating for different single-filed natural inflation models. The variation trends of reheating temperature $N_{re}$ and duration $T_{re}$ with $\omega_{re}$ and $n_s$ were explored.

The CMB constraints show that the $n_s$ and $r$ feasible space obtained by the SNI model is almost covered by the NII model, which means that the NII model is more general than the SNI model. Furthermore, the ENI model has no overlapping area with the other two models, which indicates that the ENI model and the other two models exclude each other, and more accurate experiments can separate them.

Considering the constraint of reheating, we find that the reheating equation of state $\omega_{re}$ can effectively narrow the feasible parameter space of the model, and greatly increase the accuracy of the model. Moreover, we restrict $\omega_{re}$ to the range $-\frac{1}{3}\leq\omega_{re}\leq1$, resulting in tighter constraints on the parameters of the inflation model than from the usual procedure. To this end, we explore the constraints of CMB and reheating for three modified single-filed natural inflation models and the results show that the decay constants are different for the three models, moreover, the effective equations of state $\omega_{re}$ should fall in the interval $\frac{1}{4}\lesssim \omega_{re} \lesssim \frac{4}{5}$ for three models.

\section{Acknowledgments}
This work was supported by the graduate research and innovation foundation of Chongqing, china(No.CYB21045 and No.ydstd1912), and by the China Postdoctoral Science Foundation under Grant No.2021M693743, supported by the Science Foundation of Chongqing under Grant No.D63012022005, the Fundamental Research Funds for the Central Universities under Grant No. 2020CDJQY-Z003, and the China Postdoctoral Science Foundation under Grants No. 2019TQ0329 and No. 2020M670476. The author Hua Zhou and Qing Yu thanks the financial support from China Scholarship Council.

\end{document}